\documentclass[10pt,final,journal,letter,oneside,twocolumn]{IEEEtran}

\usepackage{cite}
\usepackage{ulem}
\usepackage{soul,color}
\usepackage{srcltx}
\usepackage{multirow}
\usepackage[tbtags,sumlimits,nointlimits,reqno]{amsmath}
\usepackage{graphicx,dblfloatfix}
\graphicspath{{Figures/}}
\usepackage{float}
\usepackage{subfigure}
\usepackage{psfrag}
\usepackage{color}
\usepackage{amssymb}

\begin{document}

\title{Design of Dispersive Delay Structures (DDSs) Formed by Coupled C-Sections\\ Using Predistortion with Space Mapping}

\author{Qingfeng~Zhang,~\IEEEmembership{Member,~IEEE,}
        John~W.~Bandler,~\IEEEmembership{Life Fellow,~IEEE,}
        and~Christophe~Caloz,~\IEEEmembership{Fellow,~IEEE}
\thanks{Manuscript received July 29, 2013; revised Oct. 7, 2013; accepted Oct. 15, 2013.}
\thanks{This work was supported by NSERC Grant CRDPJ $402801$-$10$ in partnership with BlackBerry.}
\thanks{Q. Zhang and C. Caloz are with the Department of Electrical Engineering, PolyGrames Research Center,
        \'{E}cole Polytechnique de Montr\'{e}al, Montr\'{e}al, QC, Canada H3T 1J4 (email: qfzhang@ieee.org, christophe.caloz@polymtl.ca).}
\thanks{J. W. Bandler is with the Simulation Optimization Systems Research Laboratory,
Department of Electrical and Computer Engineering, McMaster University,
Hamilton, ON, Canada L8S 4K1, and also with Bandler Corporation,
Dundas, ON, Canada L9H 5E7 (e-mail: bandler@mcmaster.ca).}
\thanks{Digital Object Identifier 10.1109/TMTT.2013.2287678}
}

\markboth{IEEE TRANSACTIONS ON MICROWAVE THEORY AND TECHNIQUES}%
{Shell \MakeLowercase{\textit{et al.}}: Bare Demo of IEEEtran.cls for Journals}
\maketitle


\begin{abstract}
The concept of space mapping is applied, for the first time, to the design of microwave dispersive delay structures (DDSs). DDSs are components providing specified group delay versus frequency responses for real-time radio systems. The DDSs considered in this paper are formed by cascaded coupled C-sections. It is first shown that aggressive space mapping does not provide sufficient accuracy in the synthesis of DDSs. To address this issue, we propose a predistortion space mapping technique. Compared to aggressive space mapping, this technique provides enhanced accuracy, while compared to output space mapping, it provides greater implementation simplicity. Two full-wave and one experimental examples are provided to illustrate the proposed predistortion space mapping technique.
\end{abstract}
\IEEEoverridecommandlockouts
\begin{keywords}
Coupled C-sections, space mapping, predistortion, accuracy, dispersive delay structure (DDS), analog signal processing (ASP), real-time radio.
\end{keywords}

\IEEEpeerreviewmaketitle

\section{Introduction}

Real-time radio, a technology inspired by ultra-fast optical processing and surface acoustics wave signal processing~\cite{Lewis_SAW_OSP,SAW_ASP_Book} for high-speed microwave analog signal processing, might be a candidate to address the increasing demand for faster and more reliable wireless connectivity in the near future~\cite{caloz2013mm}. The main applications of real-time radio pertain to instrumentation~\cite{Muriel_OL_01_1999,Azana_JQE_2000,Berger_EL_2000,Laso_TMTT_03_2003,Schwartz_MWCL_04_2006,gupta2009microwave,gupta2010analog,gupta2010analog2}, radar~\cite{CRLH_DDS, Schwartz_MWCL_01_2008,CRLH_CR}, sensors~\cite{Gupta_AWPL_11_2011,Nikfal_MWCL_11_2012,qingfeng_babak_emts_2013} and communications~\cite{Yao_PTL_2000,Azana_TMTT_2007s,Nguyen_MWCL_08_2008,Xiang_TMTT_11_2012,comment_babak}. Some reviews of the real-time radio techniques are provided in~\cite{calozmetamaterial,caloz2012analog,caloz2013mm}.

The core of a real-time radio system is a dispersive delay structure (DDS), a device that controllably delays the different spectral components of an input signal by different amounts, so that the spectral information of the input signal gets mapped to the time domain of the output for real-time processing. Thus, DDSs follow group delay versus frequency specifications. They are mainly divided into reflection-type and transmission-type structures. Reflection-type DDSs can be synthesized either using Bragg grating techniques~\cite{Laso_MWCL_12_2001,Coulombe_DDS} or coupled-resonator filter techniques~\cite{0000_TMTT_Refl_Synth_Zhang}. The drawback of reflection-type DDSs is that they require circulators or hybrid couplers for transformation into two-port devices. On the other hand, transmission-type DDSs do not suffer of this issue, since they are inherently two-port devices. They can be further divided into bandpass-type and allpass-type components according to their magnitude responses. Transmission-type bandpass DDSs are designed using coupling matrix techniques~\cite{Zhang_TMTT_cross}. They are usually limited to narrow-band operation due to their resonator-based configurations. In contrast, transmission-type allpass DDSs are inherently wide-band. They are designed as coupled transmission lines~\cite{Cristal_TMTT_01_1969,steenaart1963synthesis,Gupta-allpass,Gupta_TMTT_12_2012,Gupta_IJCTA_0000,Zhang_IJRMCAE_TBP}, i.e., C-sections or D-sections. A comparison of transmission-type and reflection-type allpass DDSs in terms of system resolution was reported in~\cite{ref_trans_comparison}.

Among transmission-type allpass DDSs, \emph{uncoupled} C-section DDSs are particularly simple and can be designed using closed-form synthesis techniques~\cite{Gupta_IJCTA_0000}. However, they are relatively large. On the other hand, \emph{coupled} C-section DDSs are more compact and provide larger group delay swings, due to cross coupling. However, only highly time-consuming brute-force optimization~\cite{Gupta-allpass} has been available to design them since cross-coupling is very difficult to model accurately. Therefore, an efficient synthesis technique, benefiting from the simplicity of uncoupled C-section DDSs closed-form techniques while retaining the compactness and resolution features of coupled C-section DDSs is greatly desirable. Space mapping (SM)~\cite{bandler1995electromagnetic,bakr1998trust,bandler2002implicit,bandler2004implicit,bandler2003based} is a powerful approach to efficiently perform such a task.


We apply here SM for the first time to microwave analog signal processing, using the closed-form formulas available for uncoupled C-section DDSs for the coarse model and full-wave simulation of the corresponding coupled C-section DDS for the fine model. We start with a typical variant of SM, aggressive space mapping (ASM)~\cite{bandler1995electromagnetic}, and find that it suffers from inaccuracy in the design of DDSs due to inherent modeling differences. We subsequently propose a novel variant of SM, predistortion space mapping, which offers both enhanced accuracy and architecture simplification. In addition to providing fast convergence, this technique allows one to modify the specifications during the course of the SM iterative procedure.

The paper is organized as follows. Section~II presents the configuration of the coupled C-section DDS and its uncoupled C-section coarse model. Section~III introduces the design of coupled C-section DDS using ASM. Section~IV overviews the other variants of SM techniques and subsequently presents the proposed predistortion SM technique. Section~V provides two full-wave and one experimental examples to illustrate the proposed SM technique. Finally, conclusions are given in Sec.~VI.

\section{Coupled C-section DDS}

\begin{figure}[!t]
  \psfrag{a}[l][c]{\footnotesize strong cross coupling}
  \psfrag{k}[l][c]{\footnotesize weak cross coupling}
  \psfrag{1}[c][c]{\footnotesize $1$}
  \psfrag{2}[c][c]{\footnotesize $2$}
  \psfrag{3}[c][c]{\footnotesize $3$}
  \psfrag{4}[c][c]{\footnotesize $n$}
  \psfrag{5}[c][c]{\footnotesize $N$}
  \psfrag{l}[c][c]{\footnotesize $\ell_n$}
  \psfrag{s}[l][c]{\footnotesize $s_n$}
  \psfrag{w}[l][c]{\footnotesize $w_n$}
  \psfrag{d}[c][c]{\footnotesize $d_0$}
  \includegraphics[width=8.6cm]{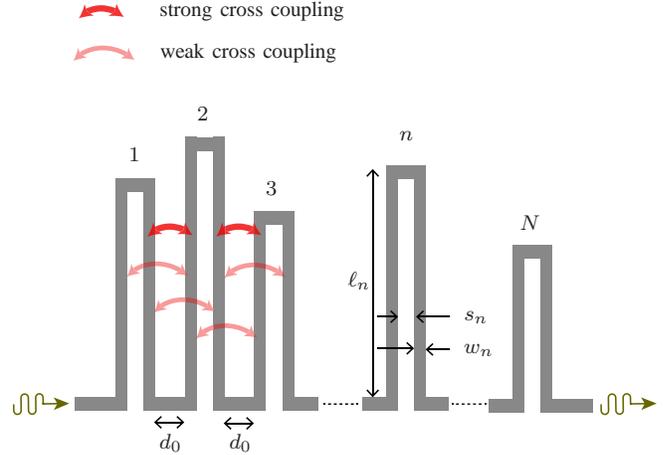}\\
  \caption{Configuration of a DDS formed by $N$ coupled C-sections (only significant cross coupling is plotted here).}\label{fig:fine_model}
\end{figure}

The coupled C-section DDS is shown in Fig.~\ref{fig:fine_model}. It features high compactness due to the small spacings between adjacent C-sections.
It is composed of $N$ coupled C-sections, which are formed by coupled transmission lines shorted at one end. The $n^\text{th}$ C-section has physical dimensions $w_n,s_n,\ell_n$, corresponding to the width, spacing and length of the coupled transmission line. The spacings between these C-sections are set here to a constant, $d_0$, which is usually set larger than the spacings of the coupled lines ($s_n$). Due to the smallness of these spacings, strong cross coupling occurs between adjacent C-sections. These couplings trap the wave inside the structure, which results in increased delays~\cite{Gupta_TMTT_12_2012,Zhang_APM_TBP}, leading to higher system resolution~\cite{caloz2013mm}.

\begin{figure}[!t]
  \psfrag{a}[c][c]{\footnotesize C-section model}
  \psfrag{1}[c][c]{\footnotesize $(\omega_{01},k_1)$}
  \psfrag{2}[c][c]{\footnotesize $(\omega_{02},k_2)$}
  \psfrag{3}[c][c]{\footnotesize $(\omega_{03},k_3)$}
  \psfrag{4}[c][c]{\footnotesize $(\omega_{0n},k_n)$}
  \psfrag{5}[c][c]{\footnotesize $(\omega_{0N},k_N)$}
  \includegraphics[width=8.6cm]{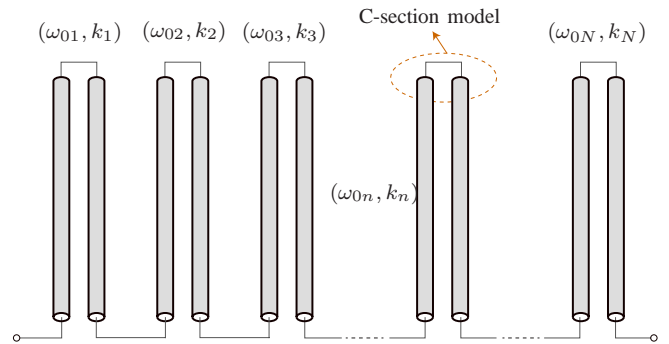}\\
  \caption{Coarse model for the DDS of Fig.~\ref{fig:fine_model}, where no cross coupling exists between the C-sections ($\omega_{0n}$ denotes the quarter-wavelength frequency and $k_n$ is the coupling factor).}\label{fig:coarse_model}
\end{figure}

It is difficult to accurately model the cross coupling between C-sections in Fig.~\ref{fig:fine_model}. However, if it is small compared to the dominant coupling associated with the individual C-sections, one may use the uncoupled structure shown in Fig.~\ref{fig:coarse_model} as the coarse model. Note that the $n^\text{th}$ C-section is characterized by the pair of parameters $(\omega_{0n},k_n)$, where  $\omega_{0n}$ denotes the frequency at which the coupled transmission line is a quarter wavelength long and $k_n$ is the coupling coefficient. The coarse model in Fig.~\ref{fig:coarse_model} can be analyzed using closed-form formulas, as follows. For the $n^\text{th}$ C-section of Fig.~\ref{fig:coarse_model}, the group delay is~\cite{steenaart1963synthesis,Gupta-allpass,Gupta_TMTT_12_2012}
  \begin{equation}\label{eq:delay_C}
    \tau_n(\omega)=\dfrac{\pi a_n}{\omega_{0n}\left[a_n^2+(1-a_n^2)\cos^2\left(\dfrac{\pi \omega}{2\omega_{0n}}\right)\right]},
  \end{equation}
  \noindent where
  \begin{equation}\label{eq:delay_a}
    a_n=\sqrt{\dfrac{1-k_n}{1+k_n}}.
  \end{equation}
Note that the group delay $\tau_n$ exhibits a periodic response, with maxima occurring at the odd multiples of $\omega_{0n}$ and minima occurring at the even multiples of $\omega_{0n}$. A wave interference explanations of the group delay dispersion in C-sections and in resonators are provided in~\cite{Gupta_TMTT_12_2012} and~\cite{Zhang_APM_TBP}, respectively.

The total group delay in uncoupled C-section DDSs of Fig.~\ref{fig:coarse_model} (coarse model) is simply the sum of the delays incurred in all the C-sections, i.e.
\begin{equation}\label{eq:delay_total}
    \tau_c(\omega)=\sum_{n=1}^N\tau_n(\omega),
\end{equation}
since no cross coupling exists between them.

Although very fast to design, this delay, based on the formula~\eqref{eq:delay_total}, is only a rough approximation of the coupled C-section DDS of interest (Fig.~\ref{fig:fine_model}) since it neglects the cross coupling effects. On the other hand, it would be very time-consuming to directly optimize the group delay of the coupled C-section DDS using full-wave analysis. ASM offers a well-suited and powerful approach to combine the speed of Fig.~\ref{fig:coarse_model} and the accuracy of Fig.~\ref{fig:fine_model} for an efficient synthesis of the DDS.

\section{DDS Optimization Using ASM}
\subsection{Statement of the Problem}

The problem is to find the parameter set $\boldsymbol{x}_f$ such that
  \begin{equation}\label{eq:fine_obj}
    \boldsymbol{R}_f(\boldsymbol{x}_f) - \boldsymbol{R}_\text{spec}=\boldsymbol{0},
  \end{equation}
\noindent where $\boldsymbol{R}_f(.)$ denotes the response of the fine model and $\boldsymbol{R}_\text{spec}$ is the specified response.
The response of the fine model ($\boldsymbol{R}_f$) corresponds to the response obtained using an accurate but time-consuming approach; in the C-section DDS problem, it will correspond to the full-wave computation of the group delay response of the coupled C-sections in Fig.~\ref{fig:fine_model}. The parameter set $\boldsymbol{x}_f$ is a vector formed by quantities that are the unknowns to find; in the C-section DDS problem, these quantities will be the physical dimensions indicated in Fig.~\ref{fig:fine_model}. $\boldsymbol{R}_f$ and $\boldsymbol{R}_\text{spec}$ are vectors whose elements represent the values of discretized functions of one or more variables at points where \eqref{eq:fine_obj} is enforced; in the C-section DDS problem, these vectors will be functions of the frequency, $\omega$, and may therefore be written explicitly $\boldsymbol{R}_f(\boldsymbol{x}_f;\boldsymbol{\omega})$ and $\boldsymbol{R}_\text{spec}(\boldsymbol{\omega})$, respectively, where $\boldsymbol{\omega}$ is a vector including the discrete frequency points.

The satisfaction of \eqref{eq:fine_obj} may be tested as
  \begin{equation}\label{eq:fine_check}
  \left\|\boldsymbol{R}_f(\boldsymbol{x}_f)-\boldsymbol{R}_\text{spec}\right\|\leq \epsilon,
  \end{equation}
\noindent where $\|.\|$ indicates a suitable norm and $\epsilon$ is a suitably small positive constant.

\subsection{ASM Review}

Instead of testing many $\boldsymbol{x}_f$'s in $\boldsymbol{R}_f(\boldsymbol{x}_f)$ to satisfy \eqref{eq:fine_check}, which is excessively time-consuming, ASM first optimizes the coarse model response $\boldsymbol{R}_c(\boldsymbol{x}_c)$ to satisfy
  \begin{equation}\label{eq:coarse_check}
  \left\|\boldsymbol{R}_c(\boldsymbol{x}_c)-\boldsymbol{R}_\text{spec}\right\|\leq \epsilon,
  \end{equation}
\noindent whose solution, $\boldsymbol{x}_c^*$, closely satisfying
\begin{equation}\label{eq:coarse_obj}
\boldsymbol{R}_c(\boldsymbol{x}_c^*) = \boldsymbol{R}_\text{spec},
\end{equation}
is almost instantaneously obtained by any classical fitting method. The coarse model response vector $\boldsymbol{R}_c$ corresponds to the response obtained using an approximate but fast approach; in the C-section DDS problem, it will correspond to the closed-form formula~\eqref{eq:delay_total} for the group delay response of Fig.~\ref{fig:coarse_model}, where the parameter set $\boldsymbol{x}_c$ will represent the quarter-wavelength frequencies and coupling coefficients in Fig.~\ref{fig:coarse_model}.

However, $\boldsymbol{x}_c^*$ is only a rough approximation to the original problem, Eq.~\eqref{eq:fine_obj}, because the coarse model is substantially different from the fine model.

In order to combine the speed of~\eqref{eq:coarse_obj} and the accuracy of~\eqref{eq:fine_obj}, one may combine the two equations via their common target $\boldsymbol{R}_\text{spec}$ as
  \begin{equation}\label{eq:new_obj}
    \boldsymbol{R}_c(\boldsymbol{x}_c^*) = \boldsymbol{R}_f(\boldsymbol{x}_f),
  \end{equation}
\noindent and then split this relation into the two new equations
  \begin{subequations}\label{eq:ASM_split_eqs_RcRfxcxcs}
  \begin{align}
    &\boldsymbol{R}_c(\boldsymbol{x}_c) = \boldsymbol{R}_f(\boldsymbol{x}_f),\label{eq:two_eq1}\\
    &\boldsymbol{x}_c-\boldsymbol{x}_c^*=\boldsymbol{0},
    \label{eq:two_eq2}
  \end{align}
  \end{subequations}
\noindent which will be solved sequentially. In this system, solving~\eqref{eq:two_eq1} is fast since this simply consists in optimizing $\boldsymbol{x}_c$ in $\boldsymbol{R}_c$ so that $\boldsymbol{R}_c(\boldsymbol{x}_c)$ fits $\boldsymbol{R}_f(\boldsymbol{x}_f)$ where $\boldsymbol{R}_f$ has been run just once. Moreover, since different $\boldsymbol{x}_f$'s clearly correspond to different $\boldsymbol{x}_c$'s in~\eqref{eq:two_eq1}, an implicit relation exists between $\boldsymbol{x}_f$ and $\boldsymbol{x}_c$. This relation may be represented by a mapping function, $\boldsymbol{P}$,
  \begin{equation}\label{eq:P}
    \boldsymbol{x}_c=\boldsymbol{P}(\boldsymbol{x}_f),
  \end{equation}
that is still undetermined at this point. So, $\boldsymbol{x}_c$ is a function of $\boldsymbol{x}_f$, and~\eqref{eq:two_eq2} is then a nonlinear equation in $\boldsymbol{x}_f$, which reads
\begin{equation}\label{eq:ASM_RcRfPxf}
\boldsymbol{P}(\boldsymbol{x}_f)-\boldsymbol{x}_c^*=\boldsymbol{0},
\end{equation}
and forms the system of equations to solve together with~\eqref{eq:two_eq1}.

To solve this system, one may introduce the error function
  \begin{equation}\label{eq:f_error}
    \boldsymbol{\eta}(\boldsymbol{x}_f)=\boldsymbol{x}_c-\boldsymbol{x}_c^*=\boldsymbol{P}(\boldsymbol{x}_f)-\boldsymbol{x}_c^*,
  \end{equation}
  \noindent with the goal
  \begin{equation}\label{eq:f_eqation}
    \boldsymbol{\eta}(\boldsymbol{x}_f)=\boldsymbol{0}.
  \end{equation}
This last relation is equivalent to the system to solve since it represents~\eqref{eq:two_eq2} via the first equality in~\eqref{eq:f_error} and~\eqref{eq:two_eq1} via the second equality in~\eqref{eq:f_error} from~\eqref{eq:ASM_RcRfPxf}.

Equation~\eqref{eq:f_eqation} may be solved by the quasi-Newton technique. Let $\boldsymbol{x}_f^{(i)}$ be its solution at the $i^\text{th}$ iteration. The next iteration updates this set as
  \begin{equation}\label{eq:iteration}
    \boldsymbol{x}_f^{(i+1)}=\boldsymbol{x}_f^{(i)}+\boldsymbol{h}^{(i)},
  \end{equation}
\noindent where $\boldsymbol{h^{(i)}}$ is obtained by inverting the relation
  \begin{equation}\label{eq:h}
    \boldsymbol{B}^{(i)}\boldsymbol{h}^{(i)}=-\boldsymbol{\eta}(\boldsymbol{x}_f^{(i)}),
  \end{equation}
\noindent where $\boldsymbol{B}^{(i)}$ is an approximation to the Jacobian matrix of the vector $\boldsymbol{\eta}$ with respect to $\boldsymbol{x}_f$ at the $i^\text{th}$ iteration. $\boldsymbol{B}^{(i)}$ is set to be the identity matrix in the first iteration ($i=1$), and is next updated using the Broyden formula~\cite{broyden1965class}
  \begin{equation}\label{eq:B}
    \boldsymbol{B}^{(i+1)}=\boldsymbol{B}^{(i)}+\dfrac{\boldsymbol{\eta}(\boldsymbol{x}_f^{(i+1)})\boldsymbol{h}^{(i)^\text{T}}}{\boldsymbol{h}^{(i)^\text{T}}\boldsymbol{h}^{(i)}}.
  \end{equation}

\subsection{Application to the Coupled C-section DDS Synthesis}

We use Fig.~\ref{fig:coarse_model} and Fig.~\ref{fig:fine_model} as the coarse and fine models, respectively. The coarse model parameter set $\boldsymbol{x}_c$ is
\begin{align}
    \boldsymbol{x}_c=[\omega_{01},\ldots,\omega_{0n},\ldots,\omega_{0N},k_1,\ldots,k_n,\ldots,k_N]^\text{T}, \label{eq:xc}
  \end{align}
\noindent where $\omega_n,k_n$ corresponds to the quarter wavelength frequency and coupling coefficient,  respectively, of the $n^\text{th}$ C-section in Fig.~\ref{fig:coarse_model}. The fine model parameter set $\boldsymbol{x}_f$ is
\begin{align}
   \boldsymbol{x}_f=[\omega'_{01},\ldots,\omega'_{0n},\ldots,\omega'_{0N},k'_1,\ldots,k'_n,\ldots,k'_N]^\text{T}, \label{eq:xf}
  \end{align}
\noindent where $\omega'_{0n},k'_n$ are the quarter wavelength frequency and coupling coefficient, respectively, of the $n^\text{th}$ C-section in Fig.~\ref{fig:fine_model}. One may relate $\{\omega'_{0n},k'_n\}$ to the physical dimensions $\{w_n,s_n,\ell_n\}$ of a C-section using approximate formula, such as given in~\cite{cohn1955shielded} for the case of a stripline implementation of Fig.~\ref{fig:fine_model}.

In practical applications of DDSs, it is the group delay swing rather than the absolute group delay matters~\cite{caloz2013mm}. Thus, group delay  responses exhibiting parallel (i.e. frequency-independent) group delay versus frequency responses are equivalent. Therefore, one may add an arbitrary constant, $\tau_0$, to the coarse model response~\eqref{eq:delay_total}, which becomes
\begin{equation}\label{eq:response_total}
    \boldsymbol{R}_c(\boldsymbol{x}_c;\boldsymbol{\omega})=\boldsymbol{\tau}_0(\boldsymbol{x}_c)+\sum_{n=1}^N\boldsymbol{\tau}_n(\omega_{0n},k_n;\boldsymbol{\omega}),
  \end{equation}
\noindent where $\boldsymbol{\tau}_n(\boldsymbol{\omega})$ is a vector whose elements are calculated by \eqref{eq:delay_C} at the corresponding elements of $\boldsymbol{\omega}$. Remember that the coarse model response in~\eqref{eq:response_total} is purely closed-form, and therefore fast to optimize, whereas the fine model response of Fig.~\ref{fig:fine_model} requires a full-wave simulation, which is accurate but excessively slow.

The key of the ASM procedure is the system~\eqref{eq:two_eq1}, which stipulates that the coarse model response must be aligned to the fine model response. However, the group delay swing of the fine model is usually larger than that of the coarse model in C-section DDSs, because of the additional cross coupling. One possible way to align the coarse model is to allow higher coupling coefficients than in the fine model so as to compensate for its cross coupling. This results in different upper bounds for the coarse and fine model, i.e.
\begin{subequations}
\begin{align}
    &k_n\in[0,U], \label{eq:bound_coarse}\\
    &k'_n\in[0,U'],\label{eq:bound_fine}
  \end{align}
\end{subequations}
\noindent where $U$ and $U'$, with $0\leq U'\leq U \leq 1$, are the upper bounds of the coarse model and fine model, respectively. $U'$ is usually determined by fabrication limitations, e.g. the minimum allowed line width and spacing. In contrast, there is no particular limitation for $U$, except $U'\leq U \leq 1$, since this quantity relates to a purely computational model. Different choices are possible for $U$, leading to different solutions to~\eqref{eq:two_eq1}, and possibly resulting in  non-converging solutions for~\eqref{eq:two_eq2}. The trust region method~\cite{bakr1998trust} is a possible approach to solve this problem. Its basic idea is to use $\boldsymbol{\eta}$ in~\eqref{eq:f_error} as a measure of the misalignment between $\boldsymbol{x}_c$ and $\boldsymbol{x}_c^*$, where this misalignment may be quantified by the geometric average error over the $2N$ elements of $\boldsymbol{\eta}$ in \eqref{eq:f_error}, i.e.
\begin{equation}\label{eq:distance}
    d=\dfrac{\sqrt{\sum_{i=1}^{2N}\eta_i^2}}{2N},
  \end{equation}
For~\eqref{eq:two_eq2} to converge, $d$ should decrease at each ASM iteration. So one should always choose the minimum $d$ among different $d$'s associated with different $U$'s.

The overall design procedure may be summarized as follows:

\begin{enumerate}
  \item Given a specified group delay response (frequency band with discretized points $\boldsymbol{\omega}$ and delay function over this band $\tau$), $\boldsymbol{R}_\text{spec}=\boldsymbol{\tau}(\boldsymbol{\omega})$, set the error in~\eqref{eq:fine_check}, $\epsilon$, to an acceptable value.

  \item Optimize the coarse model by adjusting $\boldsymbol{x}_c$ until \eqref{eq:coarse_check} is satisfied, which provides both the required order $N$ of the DDS and the aligned coarse model parameter set $\boldsymbol{x}_c^*$.

  \item Simulate the fine model with the initial setting $\boldsymbol{x}_f^{(1)}=\boldsymbol{x}_c^*$ to obtain the response $\boldsymbol{R}_f(\boldsymbol{x}_f^{(1)})$. Stop if \eqref{eq:fine_check} is satisfied. Otherwise, set $i=1$ and $\boldsymbol{B}^{(1)}=\boldsymbol{I}$, and go to the next step.

  \item Set $U'$ to the technologically highest achievable value $k'_n$ in $\boldsymbol{x}_f^{(i)}$, and find the optimal the parameter set $\boldsymbol{x}$ in the coarse model minimizing $\|\boldsymbol{R}_c(\boldsymbol{x})-\boldsymbol{R}_f(\boldsymbol{x}_f^{(i)})\|$, with a gradually increasing upper bound $U$ starting at $U=U'$ for the coupling coefficients in $\boldsymbol{x}$, until the minimum misalignment $d$ in \eqref{eq:distance} is reached. Then set $\boldsymbol{x}_c^{(i)}=\boldsymbol{x}$.

  \item Using \eqref{eq:f_error}, compute $\boldsymbol{\eta}(\boldsymbol{x}_f^{(i)})=\boldsymbol{x}_c^{(i)}-\boldsymbol{x}_c^*$, and then $\boldsymbol{h}^{(i)}$ using \eqref{eq:h}. Then update $\boldsymbol{x}_f^{(i+1)}$ using \eqref{eq:iteration}.

  \item Simulate the fine model to obtain the response $\boldsymbol{R}_f(\boldsymbol{x}_f^{(i+1)})$. Stop if \eqref{eq:fine_check} is satisfied. Otherwise, update $\boldsymbol{B}^{(i+1)}$ using \eqref{eq:B}. Set $i=i+1$ and go to Step~4).

\end{enumerate}

\subsection{Design Example}

\begin{figure}[!t]
\centering
  \psfrag{1}[c][c]{\footnotesize $1$~GHz}
  \psfrag{2}[c][c]{\footnotesize $4$~GHz}
  \psfrag{f}[c][c]{\footnotesize $f$}
  \psfrag{g}[c][c]{\footnotesize 20 mil}
  \psfrag{d}[l][c]{\footnotesize $0.5$~ns}
  \psfrag{e}[c][c]{\footnotesize $R_\text{spec}$}
  \psfrag{c}[c][c]{\footnotesize stripline}
  \psfrag{a}[c][c]{\footnotesize 0.67 mil}
  \psfrag{b}[l][c]{\footnotesize 50.67 mil}
  \includegraphics[width=7cm]{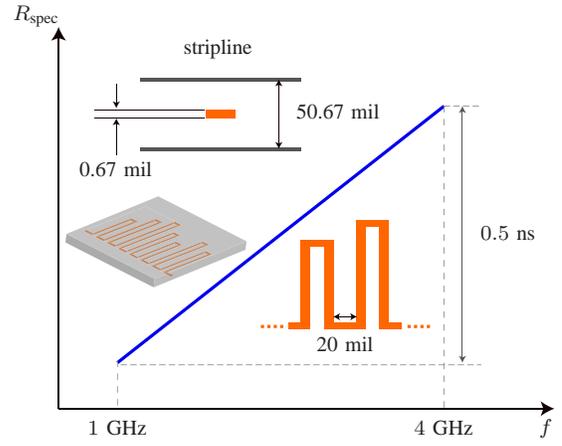}\\
  \caption{Specified group delay response for an example of a coupled C-section DDS and its configuration in stripline technology.}\label{fig:substrate}
\end{figure}

To illustrate the design method, let us take an example. Consider a specified group delay response that is linear (quadratic phase) and exhibits a swing of $0.5$~ns over the frequency band $1-4$~GHz. DDS is implemented in stripline technology, as shown in Fig.~\ref{fig:substrate}, using RT/duroid 6010LM as the substrate (dielectric constant 10.2 and loss tangent 0.0023). For convenience, all the C-sections are separated by the same gap, of $20$~mil. The physical dimensions of the C-sections in Fig.~\ref{fig:fine_model} corresponding to a parameter set $\boldsymbol{x}_f$ are calculated using the closed-form approximation formula provided in~\cite{cohn1955shielded}.

\begin{figure*}[!t]
  \center
  \subfigure[]{
  \label{fig:asm_1}
  \psfrag{a}[c][c]{\footnotesize Frequency (GHz)}
  \psfrag{b}[c][c]{\footnotesize Group delay (ns)}
  \psfrag{e}[c][c]{\footnotesize Error ($0.001$ ns)}
  \psfrag{c}[l][c]{\footnotesize $R_\text{spec}$}
  \psfrag{d}[l][c]{\footnotesize $R_c(\boldsymbol{x}_c^*)$}
  \includegraphics[width=5.8cm]{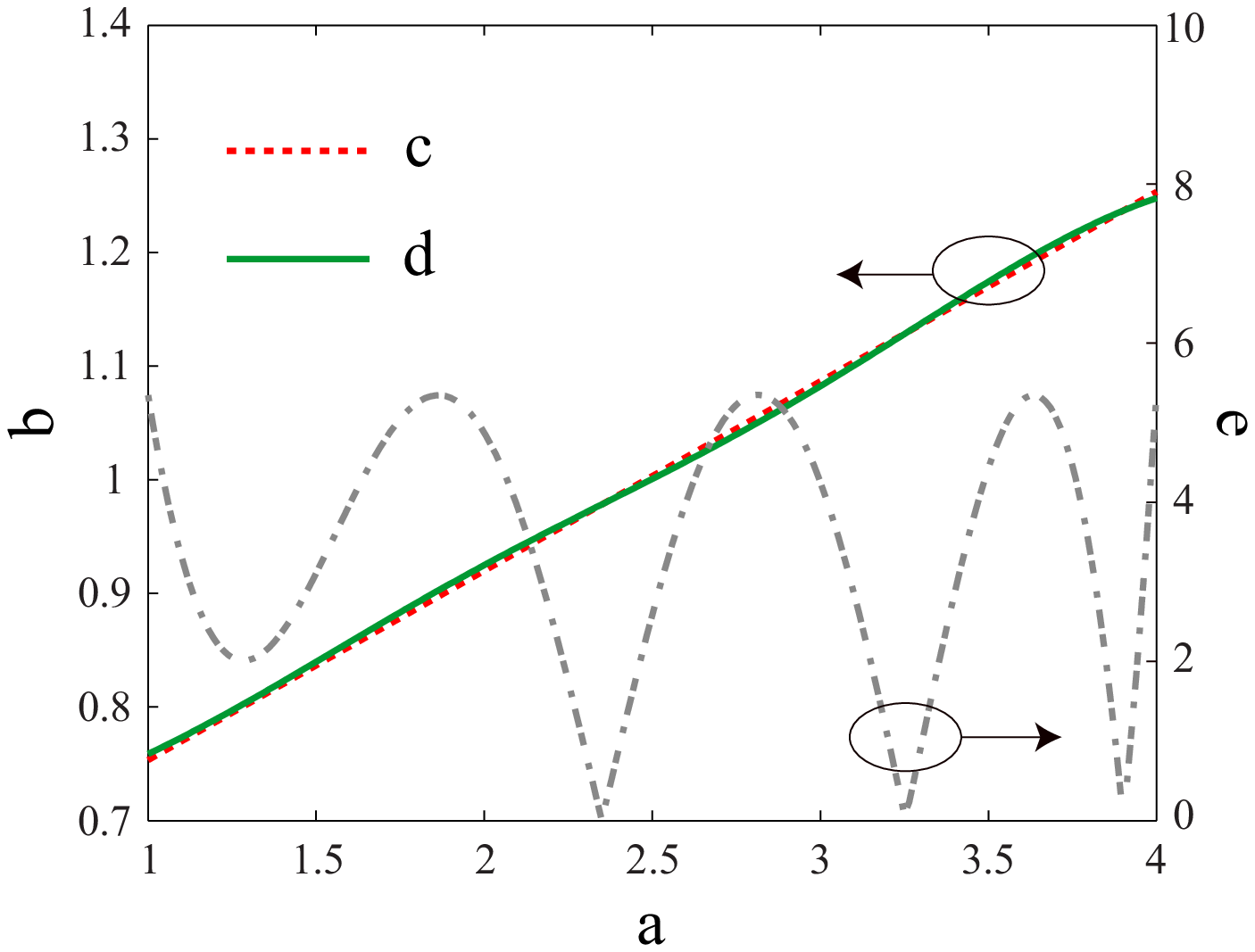}}
  \subfigure[]{
  \label{fig:asm_2}
  \psfrag{a}[c][c]{\footnotesize Frequency (GHz)}
  \psfrag{b}[c][c]{\footnotesize Group delay (ns)}
  \psfrag{e}[c][c]{\footnotesize Error ($0.1$ ns)}
  \psfrag{c}[l][c]{\footnotesize $R_\text{spec}$}
  \psfrag{d}[l][c]{\footnotesize $R_f(\boldsymbol{x}_f^{(1)})$}
  \includegraphics[width=5.8cm]{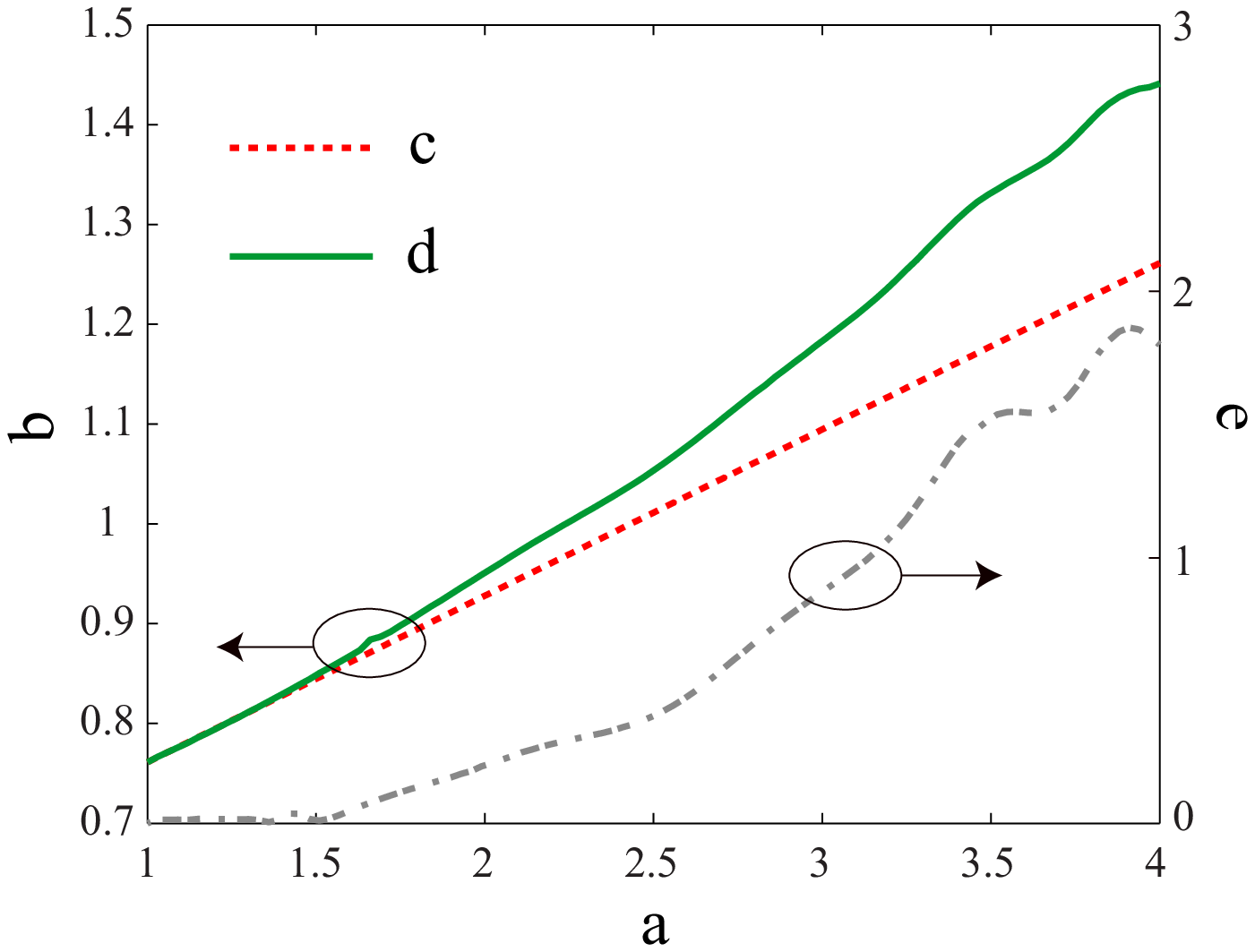}}
  \subfigure[]{
  \label{fig:asm_3}
  \psfrag{a}[c][c]{\footnotesize Frequency (GHz)}
  \psfrag{b}[c][c]{\footnotesize Group delay (ns)}
  \psfrag{e}[c][c]{\footnotesize Error ($0.01$ ns)}
  \psfrag{c}[l][c]{\footnotesize $R_\text{spec}$}
  \psfrag{d}[l][c]{\footnotesize $R_f(\boldsymbol{x}_f^{(2)})$}
  \includegraphics[width=5.8cm]{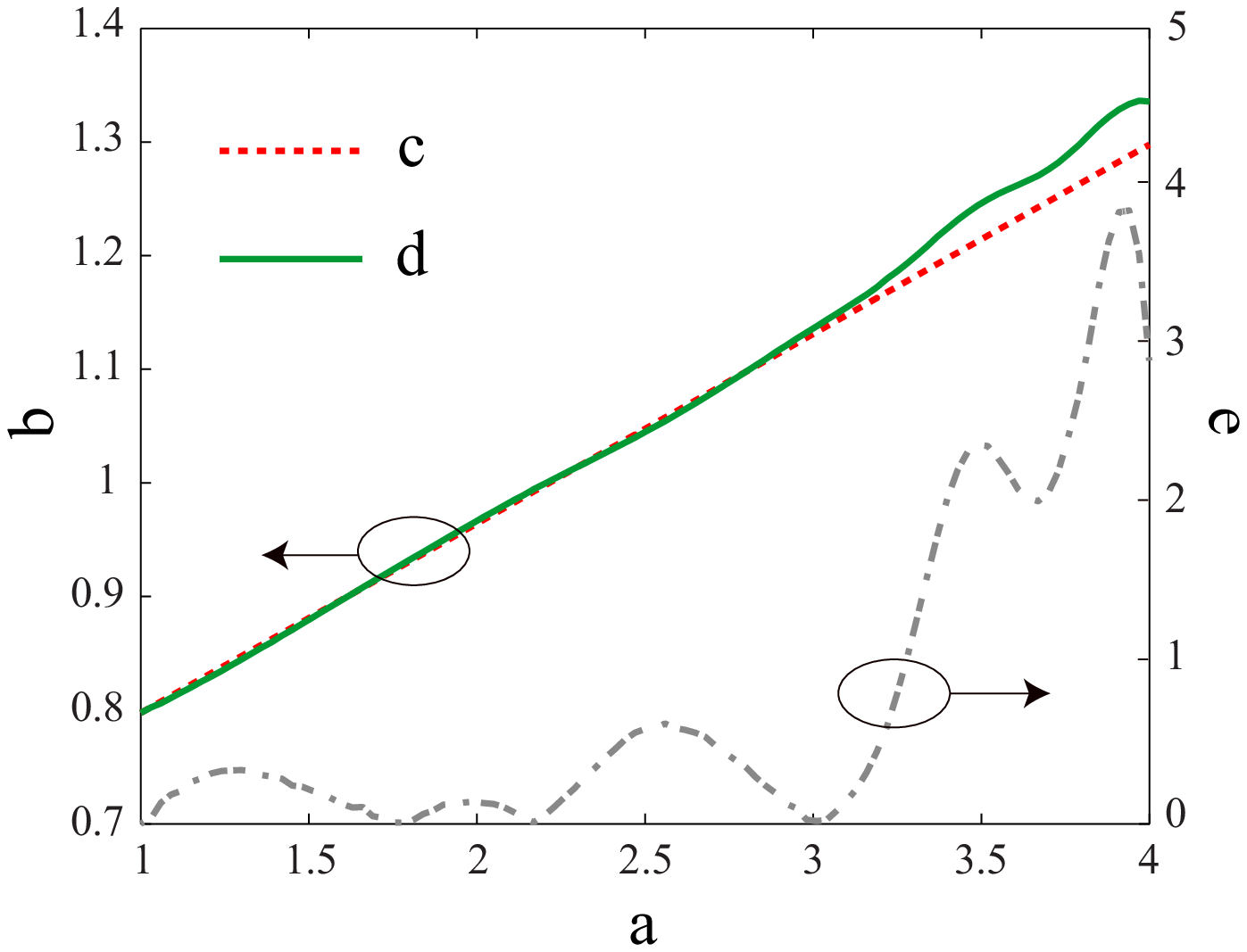}}
  \subfigure[]{
  \label{fig:asm_4}
  \psfrag{a}[c][c]{\footnotesize Frequency (GHz)}
  \psfrag{b}[c][c]{\footnotesize Group delay (ns)}
  \psfrag{e}[c][c]{\footnotesize Error ($0.01$ ns)}
  \psfrag{c}[l][c]{\footnotesize $R_\text{spec}$}
  \psfrag{d}[l][c]{\footnotesize $R_f(\boldsymbol{x}_f^{(3)})$}
  \includegraphics[width=5.8cm]{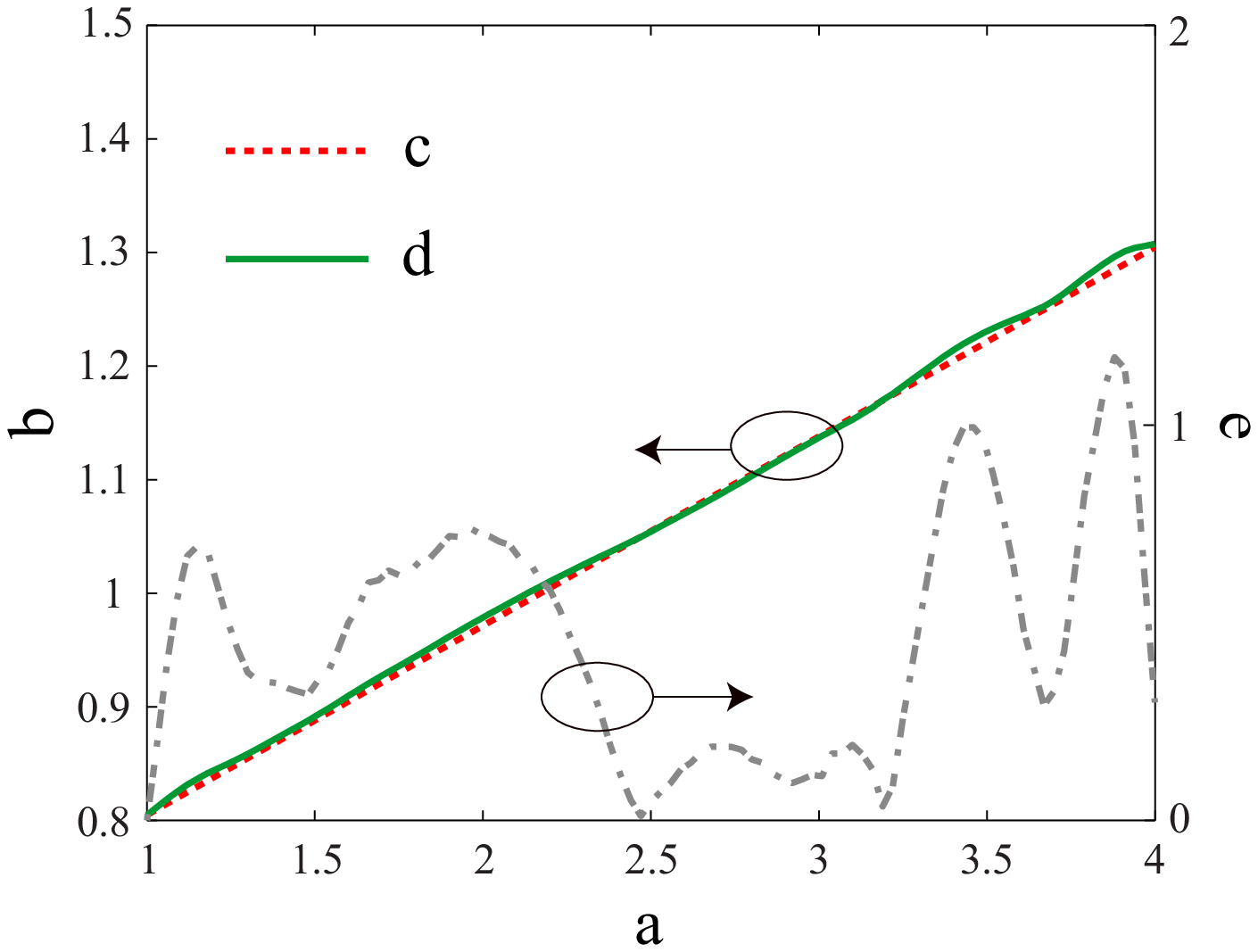}}
  \subfigure[]{
  \label{fig:asm_5}
  \psfrag{a}[c][c]{\footnotesize Frequency (GHz)}
  \psfrag{b}[c][c]{\footnotesize Group delay (ns)}
  \psfrag{e}[c][c]{\footnotesize Error ($0.001$ ns)}
  \psfrag{c}[l][c]{\footnotesize $R_\text{spec}$}
  \psfrag{d}[l][c]{\footnotesize $R_f(\boldsymbol{x}_f^{(4)})$}
  \includegraphics[width=5.8cm]{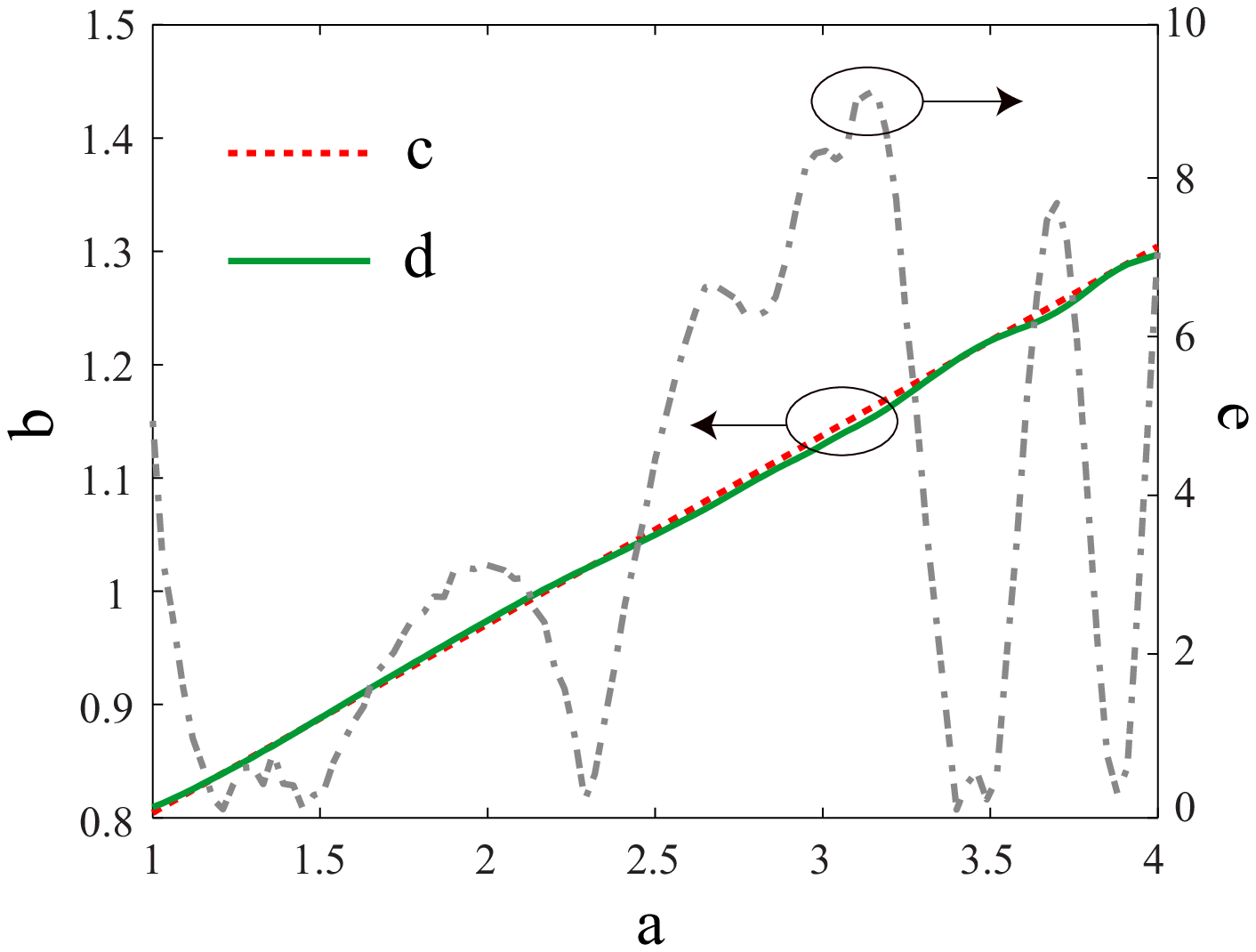}}
  \subfigure[]{
  \label{fig:asm_6}
  \psfrag{a}[c][c]{\footnotesize Frequency (GHz)}
  \psfrag{b}[c][c]{\footnotesize Group delay (ns)}
  \psfrag{e}[c][c]{\footnotesize Error ($0.001$ ns)}
  \psfrag{c}[l][c]{\footnotesize $R_\text{spec}$}
  \psfrag{d}[l][c]{\footnotesize $R_f(\boldsymbol{x}_f^{(5)})$}
  \includegraphics[width=5.8cm]{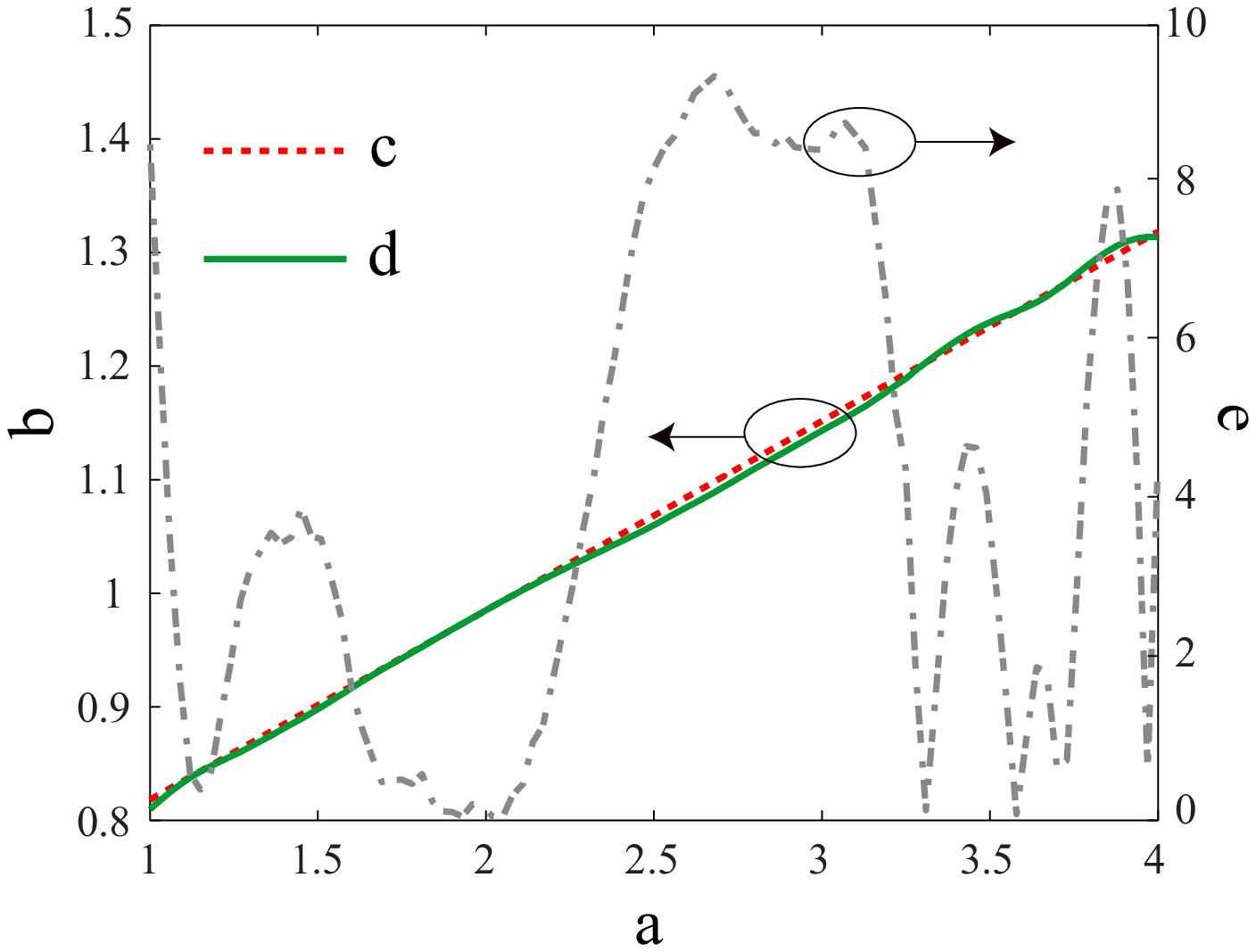}}
  \caption{Design example using ASM: group delay response of (a) the coarse model using the parameter $\boldsymbol{x}_c^*$, (b) the fine model using the parameter $\boldsymbol{x}_f^{(1)}$, (c) the fine model using the parameter $\boldsymbol{x}_f^{(2)}$, (d) the fine model using the parameter $\boldsymbol{x}_f^{(3)}$, (e) the fine model using the parameter $\boldsymbol{x}_f^{(4)}$, (f) the fine model using the parameter $\boldsymbol{x}_f^{(5)}$.}
  \label{fig:asm}
\end{figure*}

Firstly, one optimizes the coarse model parameter $\boldsymbol{x}_c^*$ to meet the specified response. The allowed maximum response error is set to $\epsilon=0.006$~ns (1.2$\%$ of the delay swing) and the subsequent number of C-sections required is found to be $N=7$. The obtained response is plotted in Fig.~\ref{fig:asm_1}. Note that the optimized response is well aligned with the specified one with an error below $\epsilon=0.006$~ns. The parameter set $\boldsymbol{x}_c^*$ is given in the first row of Tab.~\ref{tab:asm}. Note that all the C-section elements except the first one have the same parameters. This is due to the tight optimization bounds used. In this example, where $U$ is set to the relatively low value of $0.38$, most of the coupling coefficients saturate at this upper bound because of higher coupling values would actually be required to reach the relatively high specified group delay swing. This phenomenon is illustrated with different coupling bounds in the Appendix: loosely bounded parameters inherently allow for more degrees of freedom and therefore lead to a wide distribution of optimized parameter values. However, the corresponding coupling coefficients might be excessively high in practice due to fabrication limitations.

Secondly, one sets $\boldsymbol{x}_f^{(1)}=\boldsymbol{x}_c^*$ and runs the full-wave simulation. The calculated fine model response is shown in Fig.~\ref{fig:asm_2}. An expected large discrepancy, due to the cross coupling in the fine model, is observed.

\begin{table*}[!t]
\renewcommand{\arraystretch}{2}
\caption{Computed fine model parameter sets at different iterations using ASM ($f'_{0n}=\frac{\omega'_{0n}}{2\pi}/$GHz).}
\label{tab:asm}
\centering
\begin{tabular*}{1\textwidth}{@{\extracolsep{\fill}}c | c c c c c c c c c c c c c c}
\hline
\hline
$\boldsymbol{x}_f$ & $f'_{01}$ & $f'_{02}$ & $f'_{03}$ & $f'_{04}$ & $f'_{05}$ & $f'_{06}$ & $f'_{07}$ & $k'_1$ & $k'_2$ & $k'_3$ & $k'_4$ & $k'_5$ & $k'_6$ & $k'_7$\\
\hline
$\boldsymbol{x}_c^*$ & $1.957$ & $4.283$ & $4.283$ & $4.283$ & $4.283$ & $4.283$ & $4.283$ & $0.165$ & $0.380$ & $0.380$ & $0.380$ & $0.380$& $0.380$ & $0.380$\\
\hline
$\boldsymbol{x}_f^{(2)}$ & $1.934$ & $4.366$ & $4.366$ & $4.366$ & $4.366$ & $4.366$ & $4.366$ & $0.170$ & $0.290$ & $0.290$ & $0.290$ & $0.290$& $0.290$ & $0.290$\\
\hline
$\boldsymbol{x}_f^{(3)}$ & $1.967$ & $4.346$ & $4.346$ & $4.346$ & $4.346$ & $4.346$ & $4.346$ & $0.170$ & $0.257$ & $0.257$ & $0.257$ & $0.257$& $0.257$ & $0.257$\\
\hline
$\boldsymbol{x}_f^{(4)}$ & $1.991$ & $4.361$ & $4.361$ & $4.361$ & $4.361$ & $4.361$ & $4.361$ & $0.175$ & $0.253$ & $0.253$ & $0.253$ & $0.253$& $0.253$ & $0.253$\\
\hline
$\boldsymbol{x}_f^{(5)}$ & $1.945$ & $4.329$ & $4.329$ & $4.329$ & $4.329$ & $4.329$ & $4.329$ & $0.169$ & $0.257$ & $0.257$ & $0.257$ & $0.257$& $0.257$ & $0.257$\\
\hline
\hline
\end{tabular*}
\end{table*}

\begin{figure}[!t]
\centering
  \psfrag{a}[c][c]{\footnotesize $U$}
  \psfrag{b}[c][c]{\footnotesize $d$}
  \includegraphics[width=7cm]{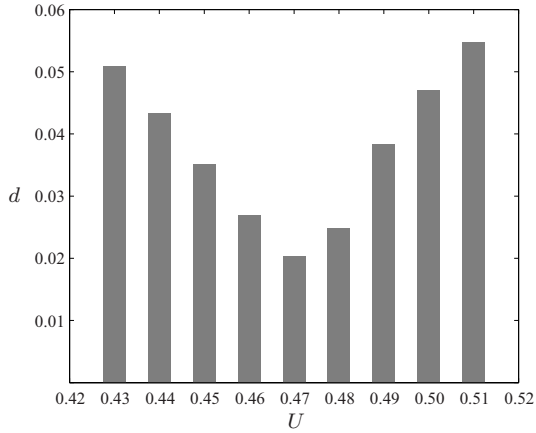}\\
  \caption{Computed misalignment $(d)$ versus the upper bound $(U)$ for the coarse model.}\label{fig:misalignment}
\end{figure}

Thirdly, one gradually increases the upper bound $U$ of the coupling coefficient $k_n$ while aligning the coarse model response with the find model distorted response. One obtains different values of $d$ for different $U$'s, as shown in Fig.~\ref{fig:misalignment}. It is noted that $d$ reaches its minimum when $U=0.47$. Therefore one chooses the optimized parameter $\boldsymbol{x}_c^{(1)}$ corresponding to $U=0.47$. Then one follows Step 5) to update $\boldsymbol{x}_f^{(2)}$, which is listed in the second row of Tab.~\ref{tab:asm}.

The response corresponding to $\boldsymbol{x}_f^{(2)}$ is shown in Fig.~\ref{fig:asm_3}. It is noted that this response is greatly improved compared with that in Fig.~\ref{fig:asm_2}. However, there is still a small discrepancy. Therefore, one runs the iteration procedure from Step 4) to Step 6) another three times. The updated parameters $\boldsymbol{x}_f^{(3)}$, $\boldsymbol{x}_f^{(4)}$ and $\boldsymbol{x}_f^{(5)}$ are listed in the third, fourth and fifth rows of Tab.~\ref{tab:asm}, respectively. The corresponding full-wave responses are plotted in Fig.~\ref{fig:asm_4}, Fig.~\ref{fig:asm_5} and Fig.~\ref{fig:asm_6}, respectively. It is observed that the responses are difficult to further improve for small discrepancies. Beyond a certain point, the error function starts to oscillate instead of converging to smaller values.

\subsection{Accuracy Issue}

It has been found in the design example that ASM technique suffers from an accuracy issue beyond some optimization point. This issue is examined next.

ASM attempts to solve \eqref{eq:coarse_obj}, \eqref{eq:two_eq1} and \eqref{eq:two_eq2} sequentially, and numerical errors inevitably incur at each of these steps and all together accumulate to produce a large final error when the variations involved are small.

One possible technique to improve accuracy is output space mapping~\cite{bandler2003based}. However, this technique involves a surrogate model in addition to the coarse model, which implies two model alignments at each iteration and therefore complicates the SM procedure. To avoid this issue, we next introduce a predistortion space mapping technique, which offers both simplicity and enhanced accuracy.

\section{Enhanced-Accuracy SM Technique}
\subsection{Overview of SM Variants}
In order to properly introduce the proposed predistortion space mapping technique, we need to briefly overview and compare the different variants of SM techniques reported to date, i.e. ASM~\cite{bandler1995electromagnetic,bakr1998trust}, implicit space mapping (ISM)~\cite{bandler2002implicit,bandler2004implicit} and output space mapping~\cite{bandler2003based}, through the perspective of equation decomposition. This will clearly show how the variants of SM techniques evolve from the original problem in~\eqref{eq:fine_obj}.

\subsubsection{ASM}
The essence of ASM is to decompose the original equation~\eqref{eq:fine_obj} into the three equations
  \begin{subequations}\label{eq:asm_sum}
  \begin{align}
    &\boldsymbol{R}_c(\boldsymbol{x}_c^*) - \boldsymbol{R}_\text{spec}=\boldsymbol{0},\\
    &\boldsymbol{R}_c(\boldsymbol{x}_c) =\boldsymbol{R}_f(\boldsymbol{x}_f),\\
    &\boldsymbol{x}_c-\boldsymbol{x}_c^*=\boldsymbol{0},
  \end{align}
  \end{subequations}
and to solve these equations sequentially, instead of solving \eqref{eq:fine_obj} directly. However, each of the three equations in \eqref{eq:asm_sum} incur numerical errors, which may accumulate and eventually lead to a large final error, as discussed in Sec.~III-E.

\subsubsection{ISM}

ISM brings in a preassigned parameter set, $\boldsymbol{x}_p$, which only appears in the coarse model. It decomposes \eqref{eq:fine_obj} into
  \begin{subequations}
  \begin{align}
    &\boldsymbol{R}_c(\boldsymbol{x}_f,\boldsymbol{x}_p) = \boldsymbol{R}_f(\boldsymbol{x}_f),\label{eq:ism_sum1}\\
    &\boldsymbol{R}_c(\boldsymbol{x}_f,\boldsymbol{x}_p) = \boldsymbol{R}_\text{spec}.\label{eq:ism_sum2}
  \end{align}\label{eq:ism_sum}
  \end{subequations}
\noindent One first optimizes $\boldsymbol{x}_p$ until \eqref{eq:ism_sum1} is satisfied, and then optimizes $\boldsymbol{x}_f$ with the optimized $\boldsymbol{x}_p$ fixed until \eqref{eq:ism_sum2} is satisfied. Note that ISM involves only two approximations compared to three approximations of ASM, and therefore it exhibits a smaller error.

\subsubsection{Output Space Mapping}

Output space mapping uses a surrogate model to improve ISM. It decomposes \eqref{eq:fine_obj} into
  \begin{subequations}
  \begin{align}
    &\boldsymbol{d} = \boldsymbol{R}_f(\boldsymbol{x}_f)-\boldsymbol{R}_c(\boldsymbol{x}_f,\boldsymbol{x}_p) ,\label{eq:osm_sum1}\\
    &\boldsymbol{R}_s(\boldsymbol{x}_f,\boldsymbol{x}_p)\stackrel{\mathrm{\Delta}}{=}\boldsymbol{R}_c(\boldsymbol{x}_f,\boldsymbol{x}_p)+\boldsymbol{d},\label{eq:osm_sum2}\\
    &\boldsymbol{R}_s(\boldsymbol{x}_f,\boldsymbol{x}_p) = \boldsymbol{R}_\text{spec},\label{eq:osm_sum3}
  \end{align}\label{eq:osm_sum}
  \end{subequations}
\noindent where $\boldsymbol{R}_s(.)$ denotes the response of the surrogate model.
To apply these equations, one first optimizes $\boldsymbol{x}_p$ to minimize the error between $\boldsymbol{R}_c(\boldsymbol{x}_f,\boldsymbol{x}_p)$ and $\boldsymbol{R}_f(\boldsymbol{x}_f)$, and then calculates the residual error using \eqref{eq:osm_sum1}. One subsequently builds the surrogate response as \eqref{eq:osm_sum2} and optimizes $\boldsymbol{x}_f$ while fixing $\boldsymbol{x}_p$ until \eqref{eq:osm_sum3} is satisfied. Since \eqref{eq:osm_sum1} and \eqref{eq:osm_sum2} are both exact, output space mapping involves only one approximation, \eqref{eq:osm_sum3}. Therefore, output space mapping would ultimately exhibit the smallest  error compared with ASM and ISM.

\subsection{Predistortion Space Mapping}

It has been pointed out in Sec. IV-A that output space mapping features a smaller error compared with ASM and ISM. However, the surrogate model induces extra complexity. In this section, we will introduce a new variant of SM techniques -- \emph{predistortion space mapping}. It is a simpler version of output space mapping featuring comparable error.

\subsubsection{Design Formula}

It has been shown in Sec. IV-A that all the variants of SM techniques follow from the original problem by equation decomposition. However, equation decomposition incurs either error accumulation or extra complexity. To avoid these issues, we shall reformulate instead of decomposing the original equation \eqref{eq:fine_obj}. One possible way to do this is to subtract $\boldsymbol{R}_c(\boldsymbol{x}_f)$ from both sides of \eqref{eq:fine_obj} and replace $\boldsymbol{x}_f$ by $\boldsymbol{x}$, which yields
  \begin{equation}\label{eq:predistortion}
    \boldsymbol{R}_c(\boldsymbol{x})=\boldsymbol{R}_c(\boldsymbol{x})-\boldsymbol{R}_f(\boldsymbol{x}) + \boldsymbol{R}_\text{spec}.
  \end{equation}
This reformulation constitutes a single equation, which avoids error accumulation and also preserves simplicity. It may be written in the iterative form
  \begin{equation}\label{eq:pre_iteration}
    \boldsymbol{R}_c(\boldsymbol{x}^{(i+1)})=\boldsymbol{R}_c(\boldsymbol{x}^{(i)})-\boldsymbol{R}_f(\boldsymbol{x}^{(i)}) + \boldsymbol{R}_\text{spec}.
  \end{equation}
\noindent Note that $\boldsymbol{R}_c(\boldsymbol{x}^{(i)})-\boldsymbol{R}_f(\boldsymbol{x}^{(i)})$ in this relation represents a response error due to the difference between the coarse and the fine models. At each iteration, one predicts this error based on the responses in the previous iteration and adds it to the specified response, which results in an updated objective response in the right-hand side of~\eqref{eq:pre_iteration}. This represents an iterative predistortion procedure since the update parameter [left-hand side of~\eqref{eq:pre_iteration}] is fitted to a distorted version of the specified function ($\boldsymbol{R}_\text{spec}$) corresponding to the sum of this function and the sum of the distance between the coarse and the fine models at the previous step of the iteration [right-hand side of~\eqref{eq:pre_iteration}].

To apply~\eqref{eq:pre_iteration}, one starts with the parameter set $\boldsymbol{x}^{(i)}$, and then optimizes the parameter set $\boldsymbol{x}^{(i+1)}$ until the condition
  \begin{equation}\label{eq:pre_stop}
    \left\|\boldsymbol{R}_c(\boldsymbol{x}^{(i+1)})-\boldsymbol{R}_c(\boldsymbol{x}^{(i)})+\boldsymbol{R}_f(\boldsymbol{x}^{(i)}) - \boldsymbol{R}_\text{spec}\right\| \leq \epsilon
  \end{equation}
\noindent is satisfied.

\subsubsection{Convergence Condition in DDS Synthesis}

Although it is very simple, the iterative formula~\eqref{eq:pre_iteration} does not necessarily converge. Therefore, one has to derive a convergence criteria for it. For this purpose, one may investigate under which condition
  \begin{equation}\label{eq:covergence_exam}
 \left\|\boldsymbol{R}_f(\boldsymbol{x}^{(i+1)}) - \boldsymbol{R}_\text{spec}\right\|<\left\|\boldsymbol{R}_f(\boldsymbol{x}^{(i)}) - \boldsymbol{R}_\text{spec}\right\|
  \end{equation}
is satisfied at each iteration, corresponding to a monotonic decrease of the error response.

Let us apply the convergence condition \eqref{eq:covergence_exam} to the synthesis of the coupled C-section DDS and try to see if it may be simplified. We start by denoting the response difference between the fine model and coarse model,
  \begin{equation}\label{eq:cross_definition}
\boldsymbol{R}_\text{cross}\stackrel{\mathrm{\Delta}}{=}\boldsymbol{R}_f-\boldsymbol{R}_c,
  \end{equation}
which represents the response contributed by cross coupling, since this is the only difference between the fine and coarse models.

Once the parameter set has been updated from $\boldsymbol{x}^{(i)}$ to $\boldsymbol{x}^{(i+1)}$, the change in the fine model response, $\delta \boldsymbol{R}_f$, differs from that in the coarse model response, $\delta \boldsymbol{R}_c$, by the amount
  \begin{equation}\label{eq:delta_analysis}
\delta \boldsymbol{R}_\text{cross}=\delta \boldsymbol{R}_f-\delta \boldsymbol{R}_c,
  \end{equation}
\noindent where
\begin{subequations}
\begin{align}
\delta \boldsymbol{R}_c&=\boldsymbol{R}_c(\boldsymbol{x}^{(i+1)})-\boldsymbol{R}_c(\boldsymbol{x}^{(i)}),\\
\delta \boldsymbol{R}_f&=\boldsymbol{R}_f(\boldsymbol{x}^{(i+1)})-\boldsymbol{R}_f(\boldsymbol{x}^{(i)}).
\end{align}\label{eq:delta_expansion}
\end{subequations}
Inserting \eqref{eq:delta_expansion} and \eqref{eq:pre_iteration} into \eqref{eq:delta_analysis} yields
\begin{align}\label{eq:delta_rf}
\delta \boldsymbol{R}_\text{cross}^{(i+1)}=\boldsymbol{R}_f(\boldsymbol{x}^{(i+1)})-\boldsymbol{R}_\text{spec},
  \end{align}
whose substitution in~\eqref{eq:covergence_exam} yields
  \begin{equation}\label{eq:covergence_exam_new}
 \left\|\delta \boldsymbol{R}_\text{cross}^{(i+1)}\right\|<\left\|\boldsymbol{R}_f(\boldsymbol{x}^{(i)}) - \boldsymbol{R}_\text{spec}\right\|.
  \end{equation}
This relation represents a sufficient condition for~\eqref{eq:pre_iteration} to converge in the coupled C-section DDS synthesis. It states that the response variation due to cross coupling at a given iteration should be smaller than the distance between the fine model response and the specified response at the previous iteration.

At the beginning of the iterative procedure, the fine model response is usually far from the specified one. Then~\eqref{eq:covergence_exam_new} is easy to satisfy since its right-hand side is large while the left-hand part is small. Accordingly, the iteration procedure at the beginning is inherently convergent without any condition.

\begin{figure}[!t]
\centering
  \psfrag{1}[c][c]{\footnotesize $\omega_1$}
  \psfrag{2}[c][c]{\footnotesize $\omega_2$}
  \psfrag{f}[c][c]{\footnotesize $\omega$}
  \psfrag{e}[c][c]{\footnotesize $\tau$}
  \psfrag{4}[l][c]{\footnotesize $R_\text{spec}$}
  \psfrag{3}[l][c]{\footnotesize $R_c(\boldsymbol{x}^{(1)})$}
  \psfrag{a}[r][c]{\footnotesize $R_f(\boldsymbol{x}^{(1)})$}
  \psfrag{b}[c][c]{\footnotesize $R_c(\boldsymbol{x}^{(2)})$}
  \psfrag{c}[l][c]{\footnotesize \shortstack{possible region\\for $R_f(\boldsymbol{x}^{(2)})$}}
  \includegraphics[width=7cm]{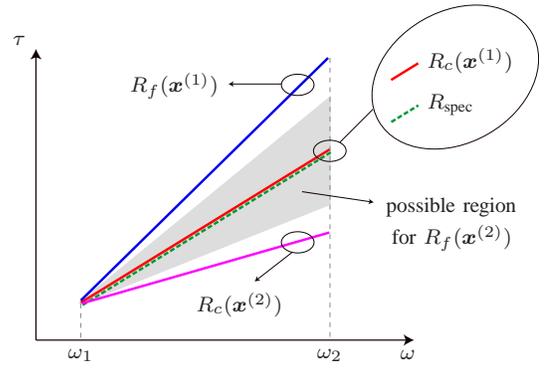}\\
  \caption{Illustration of the convergence behavior of predistorsion space mapping after the first iteration (the group delay responses are all regarded as linear for simplicity).}\label{fig:iteration_beginning}
\end{figure}

To illustrate this point, let us consider the first two iterations for a simple design example, as depicted in Fig.~\ref{fig:iteration_beginning}. Assume that the specified group delay is linear, as represented by the dotted green line in Fig.~\ref{fig:iteration_beginning}. One first optimizes the parameter set $\boldsymbol{x}^{(1)}$ until the coarse model response, plotted in solid red in Fig.~\ref{fig:iteration_beginning}, aligns with the specified one. Next $\boldsymbol{x}^{(1)}$ is used to obtain the fine model response\footnote{Here, all the responses are regarded as linear for simplicity. In practice, they are imperfectly linear, even for a linear group delay specification, as seen for instance in the green curve of Fig.~\ref{fig:asm_2}.}, plotted in solid blue in Fig.~\ref{fig:iteration_beginning}. Note that the blue line has a larger slope than the red line due to the extra cross coupling in the fine model. One next optimizes a new parameter set $\boldsymbol{x}^{(2)}$ until \eqref{eq:pre_iteration} is satisfied. The corresponding coarse model response, $R_c(\boldsymbol{x}^{(2)})$, is plotted in solid magenta in Fig.~\ref{fig:iteration_beginning}. Where will then $R_f(\boldsymbol{x}^{(2)})$ be located with respect to the other curves? Firstly, it should be below the $\boldsymbol{x}^{(1)}$ (blue) curve since the response of the coarse model is decreasing (from the red curve to the magenta curve). Secondly, it should be above the purple line because cross coupling, only existing in the fine model, increases the group delay swing. Therefore, the possible region for $R_f(\boldsymbol{x}^{(2)})$ is the shaded area in Fig.~\ref{fig:iteration_beginning}. It is then obvious that \eqref{eq:covergence_exam} is satisfied, if one considers the integral area under the curve as the norm for~\eqref{eq:covergence_exam}.

However, when the fine model response has become close to the specified one, after a few iterations, Equation~\eqref{eq:covergence_exam_new} may not be satisfied any more, due to the small value of its right-hand side. In this case, one has to enforce a special condition to maintain convergence. One possible way is to do so is to prescribe
  \begin{equation}\label{eq:cross_zero}
\delta \boldsymbol{R}_\text{cross}\approx \boldsymbol{0},
  \end{equation}
which will naturally maintain convergence in \eqref{eq:covergence_exam_new} since its left-hand side is now forced to be almost zero.

How can the condition \eqref{eq:cross_zero} be integrated in the DDS synthesis? Equation~\eqref{eq:cross_zero} states that cross coupling is forbidden to vary significantly when the parameter set $\boldsymbol{x}$ is updated. One way to realize this is to fix in $\boldsymbol{x}$ the parameters that are most sensitive to cross coupling.

\begin{figure}[!t]
\centering
  \psfrag{1}[c][c]{\footnotesize $\{w_1,s_1,\ell_1\}$}
    \psfrag{2}[c][c]{\footnotesize $\{w_2,s_2,\ell_2\}$}
      \psfrag{3}[c][c]{\footnotesize $\{w_3,s_3,\ell_3\}$}
  \psfrag{d}[c][c]{\footnotesize $d_0$}
    \psfrag{a}[c][c]{\footnotesize cross coupling}
  \includegraphics[width=5cm]{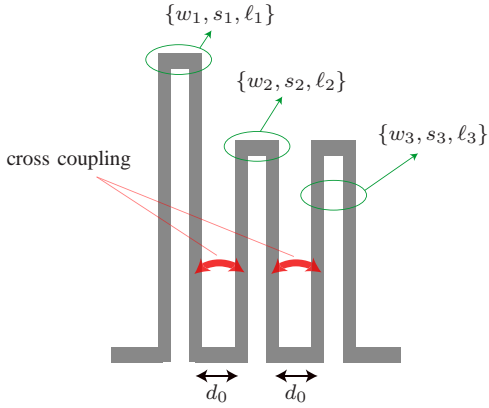}\\
  \caption{DDS example formed by three coupled C-sections. The parameters $\{w_n,s_n,\ell_n\}$ for the $n^\text{th}$ C-section are defined in Fig.~\ref{fig:fine_model}.}\label{fig:converge_condition}
\end{figure}

To illustrate this strategy, let us consider a simple DDS formed by three C-sections, as shown in Fig.~\ref{fig:converge_condition}. Cross coupling with respect to the first C-section is very sensitive to the spacing ($d_0$) between the C-sections and to the lengths of the second and third C-sections, $\ell_2$ and $\ell_3$, respectively. So, one may fix the three parameters while optimizing the other parameters in $\boldsymbol{x}$, so that the response contributed by cross coupling remains almost unchanged.

In the problem of the C-section DDS, the proposed predistortion space mapping technique automatically converges if condition~(33) is satisfied, irrespectively of the magnitude of the initial distance between the fine model response and the specification. Therefore, it can be applied independently, without prior resorting to ASM, from the beginning to the end of the optimization procedure. This is true in the particular case of the C-section DDS, as shown above, but may not be true in general. In other problems, ASM is still recommended as a first phase of the optimization procedure, until the misalignment between the model response and the specification is sufficiently small for predistorsion space mapping to converge.

\subsubsection{Design Procedure}

The converging predistorsion SM procedure for the coupled C-section DDS may be summarized as follows:

\begin{enumerate}
  \item[a)] Given a specified group delay response, $\boldsymbol{R}_\text{spec}$, choose $\epsilon$.

  \item[b)] Optimize the coarse model until it satisfies~\eqref{eq:coarse_check}, to obtain the required order $N$ and the parameter set $\boldsymbol{x}_c^*$.

  \item[c)] Set $\boldsymbol{x}^{(1)}=\boldsymbol{x}_c^*$. Simulate the fine model and obtain the response $\boldsymbol{R}_f(\boldsymbol{x}^{(1)})$. Stop if \eqref{eq:fine_check} is satisfied. Otherwise, set $i=1$ and go to Step~d).

  \item[d)] Optimize the coarse model with the parameter set $\boldsymbol{x}^{(i+1)}$ until \eqref{eq:pre_stop} is satisfied.

  \item[e)] Simulate the fine model to obtain the response $\boldsymbol{R}_f(\boldsymbol{x}^{(i+1)})$. Stop if \eqref{eq:fine_check} is satisfied. Otherwise, go to Step~f).

  \item[f)] Test if the condition \eqref{eq:covergence_exam} is satisfied. If yes, set $i=i+1$ and go to Step~d). Otherwise, go to Step g).

  \item[g)] Fix in $\boldsymbol{x}^{(i+1)}$ the parameters that are related to the cross coupling and optimize the other parameters until \eqref{eq:pre_stop} is satisfied.

  \item[h)] Simulate the fine model to obtain the response $\boldsymbol{R}_f(\boldsymbol{x}^{(i+1)})$. Stop if \eqref{eq:fine_check} is satisfied. Otherwise, set $i=i+1$ and go to Step~g).

\end{enumerate}

\subsection{Comparison of SM Variants}

Table~\ref{tab:variants} compares different SM variants for the DDS design. Aggressive SM and predistortion SM provides better initial improvement, while implicit SM and output SM offer more flexibility since they allow choices for the implicit parameter (as $x_p$ in \eqref{eq:ism_sum} and \eqref{eq:osm_sum}). In terms of simplicity, predistortion SM, using only one equation as in \eqref{eq:predistortion}, is much simpler than aggressive SM involving three equations as in \eqref{eq:asm_sum}. For final accuracy, predistortion SM provides better results than aggressive SM, as can be seen from Fig.~\ref{fig:asm} and Fig.~\ref{fig:pre} in the forthcoming section. For the initial coarse model optimization, it is automatically included in aggressive SM, whereas it is preferable for the other three SM variants.

\begin{table*}[!t]
\renewcommand{\arraystretch}{1.5}
\caption{Comparison of different SM variants for the DDS design.}
\label{tab:variants}
\centering
\begin{tabular*}{0.85\textwidth}{@{\extracolsep{\fill}}l | c | c | c | c | c}
\hline
\hline
SM Variants & Initial Improvement & Flexibility & Simplicity & Final Accuracy & Initial Coarse Model Optimization\\
\hline
\hline
Aggressive SM & High & Less & Less & Less & Included\\
\hline
Implicit SM & Moderate & High & Moderate & Moderate & Preferable\\
\hline
Output + Implicit SM & Moderate & High & Less & High & Preferable\\
\hline
Predistortion SM & High & Less & High & High & Preferable\\
\hline
\hline
\end{tabular*}
\end{table*}

\section{Predistortion Space Mapping Design Examples}

\subsection{Numerical (Full-Wave) Examples}

To illustrate the proposed predistortion space mapping technique, we apply it to the design example addressed with ASM in Sec. III-D. The specified group delay is still linear with a swing of $0.5$~ns over the frequency band $1-4$~GHz. The maximum allowed response error $\epsilon$ is $0.006$~ns.

\begin{figure*}[!t]
  \center
  \subfigure[]{
  \label{fig:pre_1}
  \psfrag{a}[c][c]{\footnotesize Frequency (GHz)}
  \psfrag{b}[c][c]{\footnotesize Group delay (ns)}
  \psfrag{e}[c][c]{\footnotesize Error ($0.1$ ns)}
  \psfrag{c}[l][c]{\footnotesize $R_\text{spec}$}
  \psfrag{d}[l][c]{\footnotesize $R_f(\boldsymbol{x}^{(1)})$}
  \includegraphics[width=5.8cm]{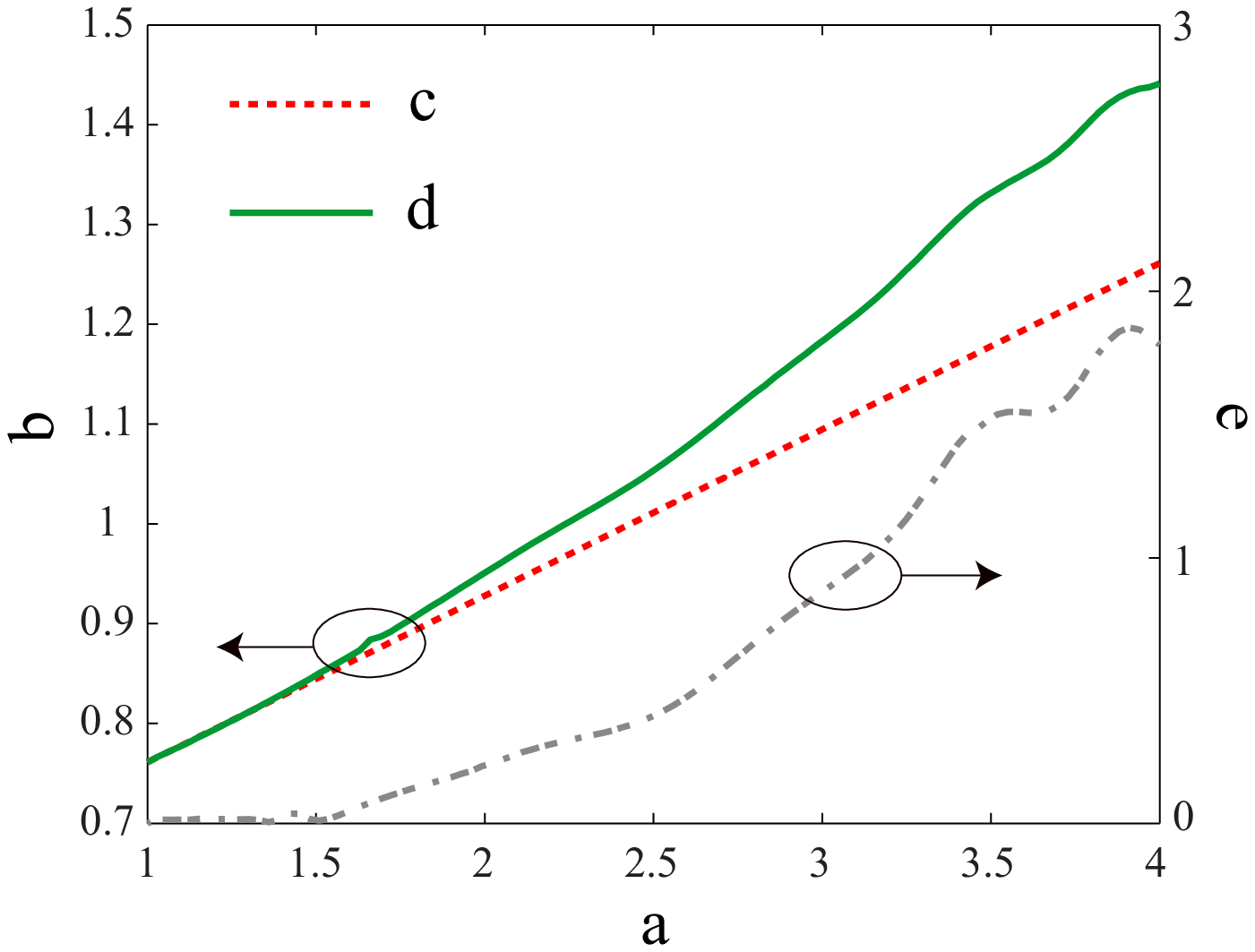}}
  \subfigure[]{
  \label{fig:pre_2}
  \psfrag{a}[c][c]{\footnotesize Frequency (GHz)}
  \psfrag{b}[c][c]{\footnotesize Group delay (ns)}
  \psfrag{e}[c][c]{\footnotesize Error ($0.001$ ns)}
  \psfrag{c}[l][c]{\footnotesize $R_\text{spec}$}
  \psfrag{d}[l][c]{\footnotesize $R_f(\boldsymbol{x}^{(2)})$}
  \includegraphics[width=5.8cm]{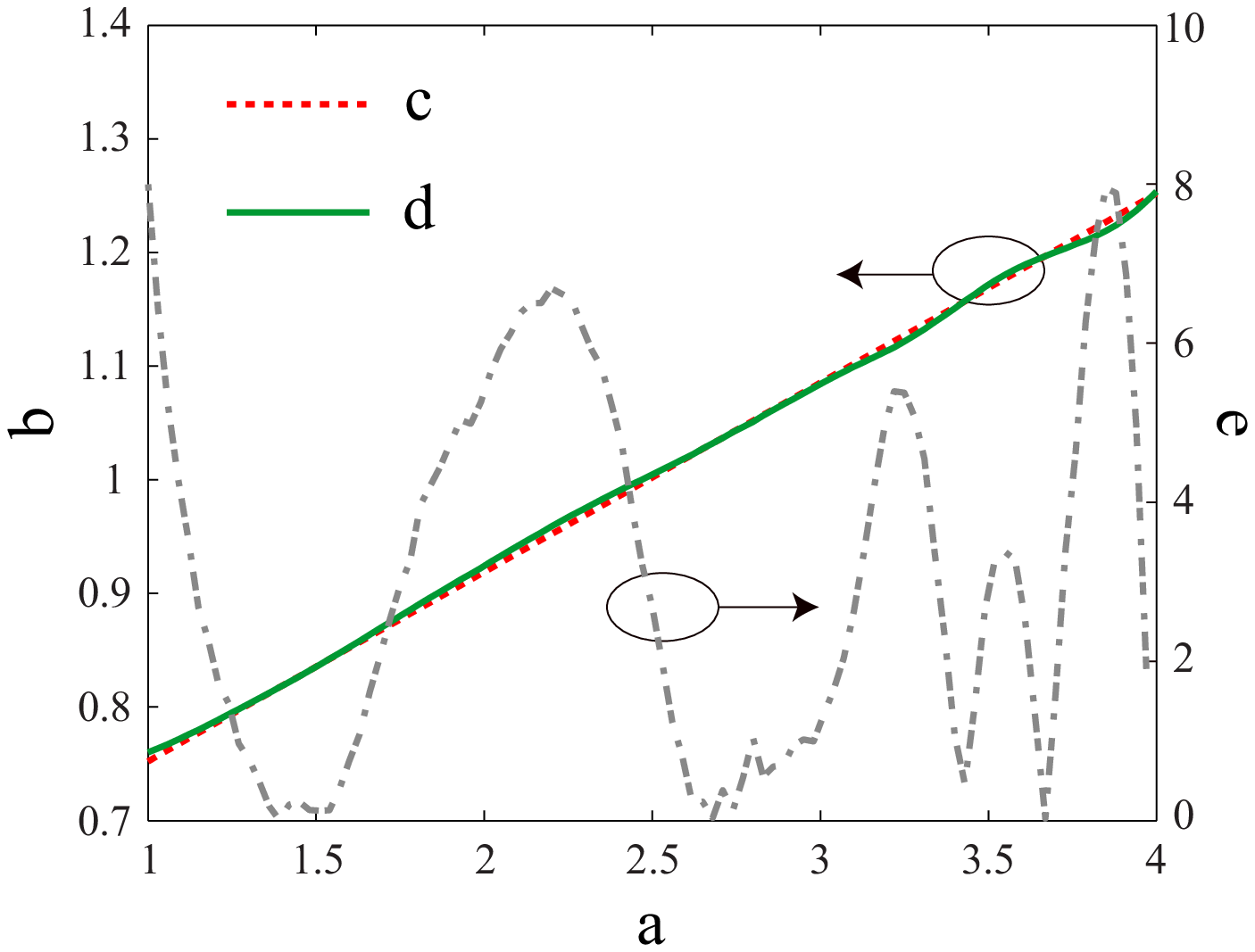}}
  \subfigure[]{
  \label{fig:pre_3}
  \psfrag{a}[c][c]{\footnotesize Frequency (GHz)}
  \psfrag{b}[c][c]{\footnotesize Group delay (ns)}
  \psfrag{e}[c][c]{\footnotesize Error ($0.001$ ns)}
  \psfrag{c}[l][c]{\footnotesize $R_\text{spec}$}
  \psfrag{d}[l][c]{\footnotesize $R_f(\boldsymbol{x}^{(3)})$}
  \includegraphics[width=5.8cm]{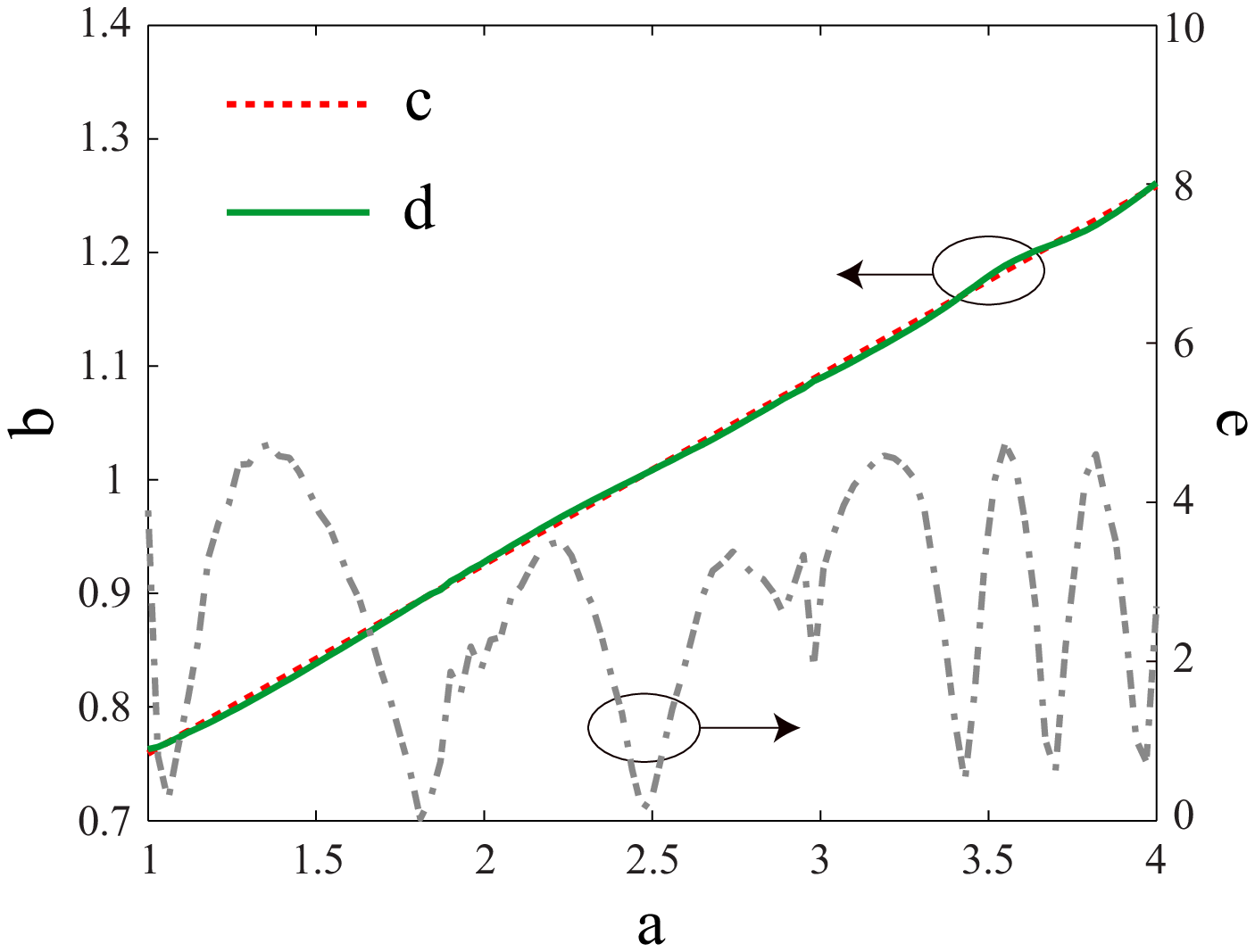}}
  \caption{Design example using the proposed predistortion space mapping technique: group delay response of the fine model using the parameter set (a)~$\boldsymbol{x}^{(1)}$, (b)~$\boldsymbol{x}^{(2)}$, and (c)~$\boldsymbol{x}^{(3)}$.}
  \label{fig:pre}
\end{figure*}

The first step of the predistortion space mapping technique is exactly the same as that of ASM. One aligns the coarse model to the specified response and obtain the parameter set $\boldsymbol{x}_c^*$. The synthesized coarse model response is plotted in Fig.~\ref{fig:asm_1}. Then one sets $\boldsymbol{x}^{(1)}=\boldsymbol{x}_c^*$ and runs a full-wave simulator for the fine model. The computed response is plotted in Fig.~\ref{fig:pre_1}, and the parameter set $\boldsymbol{x}^{(1)}$ is given in the first row of Tab.~\ref{tab:pre}. Then one runs step d) and e) in the design procedure outlined in Sec.~IV-B. The updated parameter set $\boldsymbol{x}^{(2)}$ is given in the second row of Tab.~\ref{tab:pre}, and the fine model response is plotted in Fig.~\ref{fig:pre_2}. Comparing this response with that in Fig.~\ref{fig:asm_3} confirms that the predistortion space mapping technique converges much faster than ASM. Also note that the error is still above the allowed level. Therefore, one runs another round of correction through steps d)-f). The updated parameter set $\boldsymbol{x}^{(2)}$ is given in the third row of Tab.~\ref{tab:pre}, and the calculated fine model response is plotted in Fig.~\ref{fig:pre_3}. Note that the response closely follows the specified one and that the error is below the allowed level. Accordingly, the predistortion space mapping is proved to provide enhanced convergence in both speed and accuracy.

\begin{table*}[!t]
\renewcommand{\arraystretch}{2}
\caption{Computed fine model parameter sets at different iterations using the predistortion space mapping technique ($f'_{0n}=\frac{\omega'_{0n}}{2\pi}/$GHz).}
\label{tab:pre}
\centering
\begin{tabular*}{1\textwidth}{@{\extracolsep{\fill}}c | c c c c c c c c c c c c c c}
\hline
\hline
$\boldsymbol{x}$ & $f'_{01}$ & $f'_{02}$ & $f'_{03}$ & $f'_{04}$ & $f'_{05}$ & $f'_{06}$ & $f'_{07}$ & $k'_1$ & $k'_2$ & $k'_3$ & $k'_4$ & $k'_5$ & $k'_6$ & $k'_7$\\
\hline
$\boldsymbol{x}^{(1)}$ & $1.957$ & $4.283$ & $4.283$ & $4.283$ & $4.283$ & $4.283$ & $4.283$ & $0.165$ & $0.380$ & $0.380$ & $0.380$ & $0.380$& $0.380$ & $0.380$\\
\hline
$\boldsymbol{x}^{(2)}$ & $2.134$ & $4.499$ & $4.499$ & $4.499$ & $4.499$ & $4.499$ & $4.499$ & $0.197$ & $0.270$ & $0.270$ & $0.270$ & $0.270$& $0.270$ & $0.270$\\
\hline
$\boldsymbol{x}^{(3)}$ & $2.117$ & $4.485$ & $4.485$ & $4.485$ & $4.485$ & $4.485$ & $4.485$ & $0.197$ & $0.275$ & $0.275$ & $0.275$ & $0.275$& $0.275$ & $0.275$\\
\hline
\hline
\end{tabular*}
\end{table*}

Another benefit of the predistortion space mapping technique is that it allows one to alter the specification, if desirable, at any iteration of the design procedure, without having to start it over.  In practical analog signal processing applications, it is typically desired to achieve the highest possible group delay swing. Assume therefore that the specification is given in terms of a \emph{minimum} group delay swing,  $\Delta\tau_\text{spec}^\text{min}$. Further assume that there is a specified limitation in terms of the size of the DDS component, which allows no more than $N^\text{max}$ cascaded C-sections. In this case, the procedure is launched with $N=N^\text{max}$ and the initial swing $\Delta\tau_\text{spec}^\text{I}=\Delta\tau_\text{spec}^\text{min}$. Moreover, the maximal acceptable coupling coefficient is set,  based on fabrication constraints, to $k_\text{fab}^\text{max}$, where it is noted that the maximal coupling coefficient required to meet $\Delta\tau_\text{spec}^\text{I}$, i.e. the greatest $k$ among all the $k$'s of the C-sections of the DDS, $k^\text{max}$ might not need to be as large as $k_\text{fab}^\text{max}$\footnote{The value of $k^\text{max}$  that the procedure will find to meet the specifications is unknown at the beginning of the procedure since cross coupling contributions are still unknown at this stage.}. If at a given iteration one finds that $k^\text{max}$ is much smaller than $k_\text{fab}^\text{max}$, this means that there is margin in the $k$'s to achieve a substantially greater swing than $\Delta\tau_\text{spec}^\text{min}$. Therefore, one may change on the fly the swing specification to $\Delta\tau_\text{spec}^\text{II}>\Delta\tau_\text{spec}^\text{min}$ for the next iteration. If at that point the new $k^\text{max}$ is still smaller than $k_\text{fab}^\text{max}$, but close to it, then the maximal swing that could be obtained for the allowed number of C-section has been reached, and one keeps going with the procedure; otherwise, if  $k^\text{max}$ is greater than $k_\text{fab}^\text{max}$, then $\Delta\tau_\text{spec}^\text{II}$ was too ambitious, and one needs to target a smaller swing,  $\Delta\tau_\text{spec}^\text{III}$, such that $\Delta\tau_\text{spec}^\text{I}<\Delta\tau_\text{spec}^\text{III}<\Delta\tau_\text{spec}^\text{II}$.

\begin{figure*}[!t]
  \center
  \subfigure[]{
  \label{fig:mid_1}
  \psfrag{a}[c][c]{\footnotesize Frequency (GHz)}
  \psfrag{b}[c][c]{\footnotesize Group delay (ns)}
  \psfrag{e}[c][c]{\footnotesize Error ($0.01$ ns)}
  \psfrag{c}[l][c]{\footnotesize Original $R_\text{spec}$}
  \psfrag{f}[l][c]{\footnotesize New $R_\text{spec}$}
  \psfrag{d}[l][c]{\footnotesize $R_f(\boldsymbol{x}^{(2)})$}
  \includegraphics[width=5.8cm]{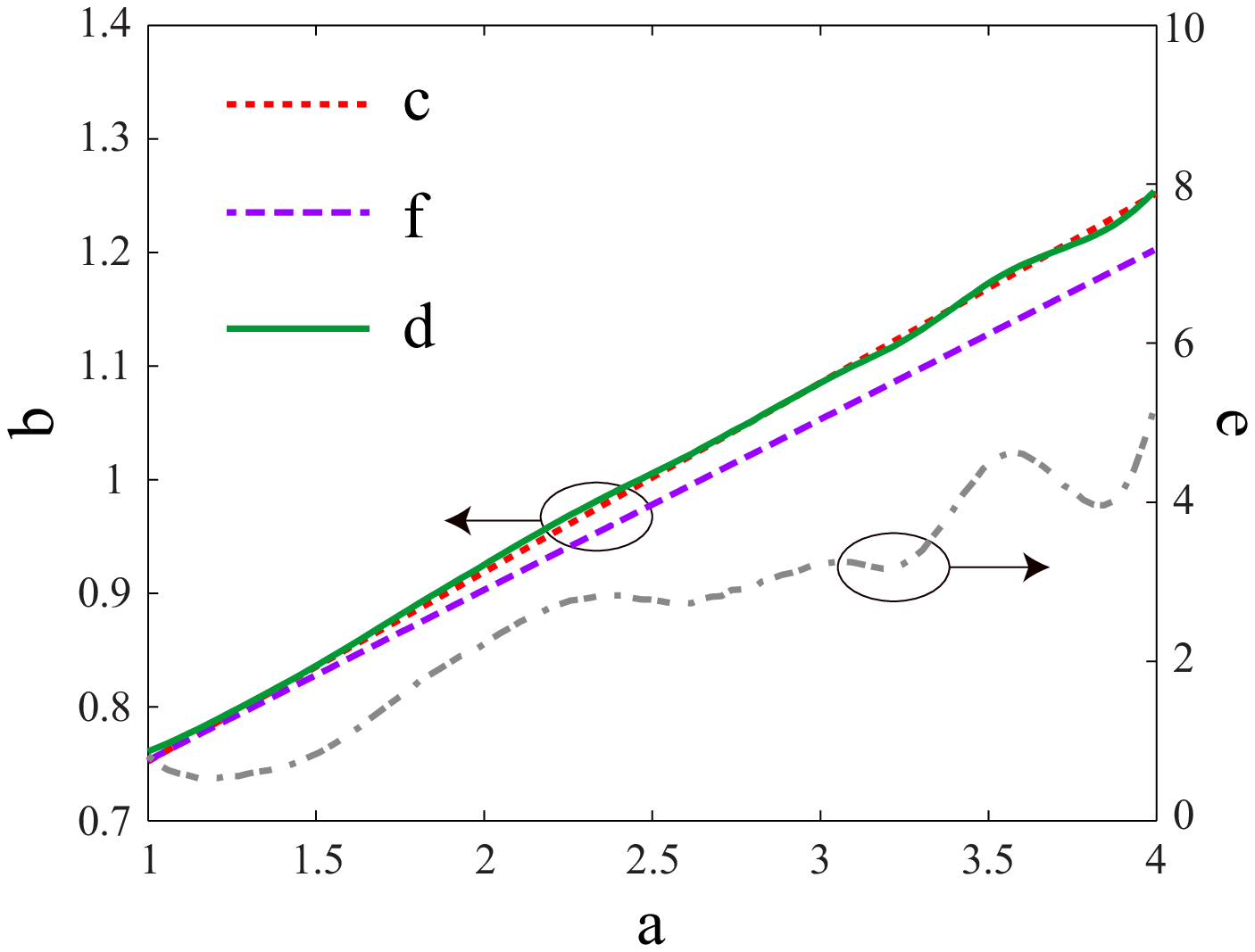}}
  \subfigure[]{
  \label{fig:mid_2}
  \psfrag{a}[c][c]{\footnotesize Frequency (GHz)}
  \psfrag{b}[c][c]{\footnotesize Group delay (ns)}
  \psfrag{e}[c][c]{\footnotesize Error ($0.01$ ns)}
  \psfrag{c}[l][c]{\footnotesize New $R_\text{spec}$}
  \psfrag{d}[l][c]{\footnotesize $R_f(\boldsymbol{x}^{(3)})$}
  \includegraphics[width=5.8cm]{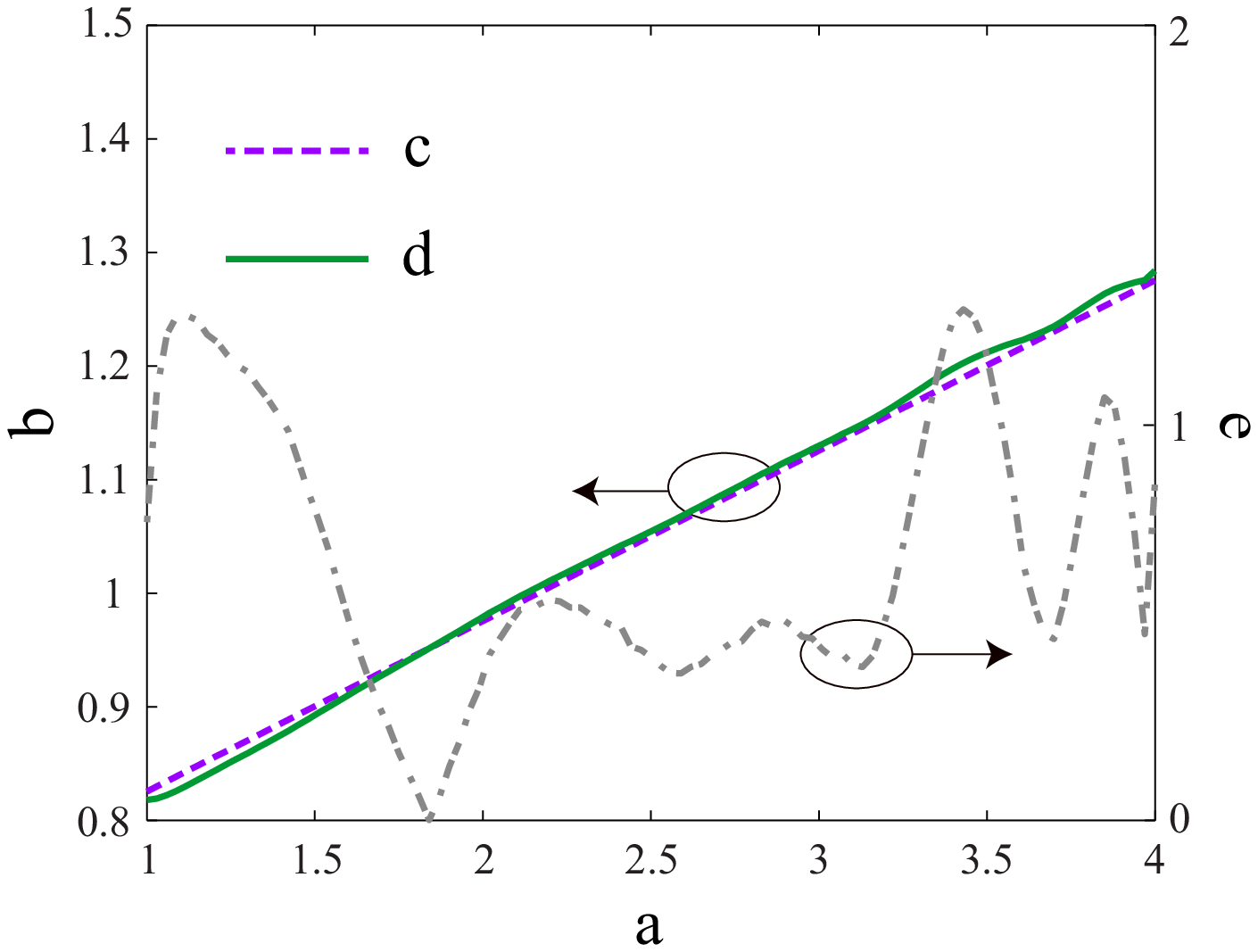}}
  \subfigure[]{
  \label{fig:mid_3}
  \psfrag{a}[c][c]{\footnotesize Frequency (GHz)}
  \psfrag{b}[c][c]{\footnotesize Group delay (ns)}
  \psfrag{e}[c][c]{\footnotesize Error ($0.001$ ns)}
  \psfrag{c}[l][c]{\footnotesize New $R_\text{spec}$}
  \psfrag{d}[l][c]{\footnotesize $R_f(\boldsymbol{x}^{(4)})$}
  \includegraphics[width=5.8cm]{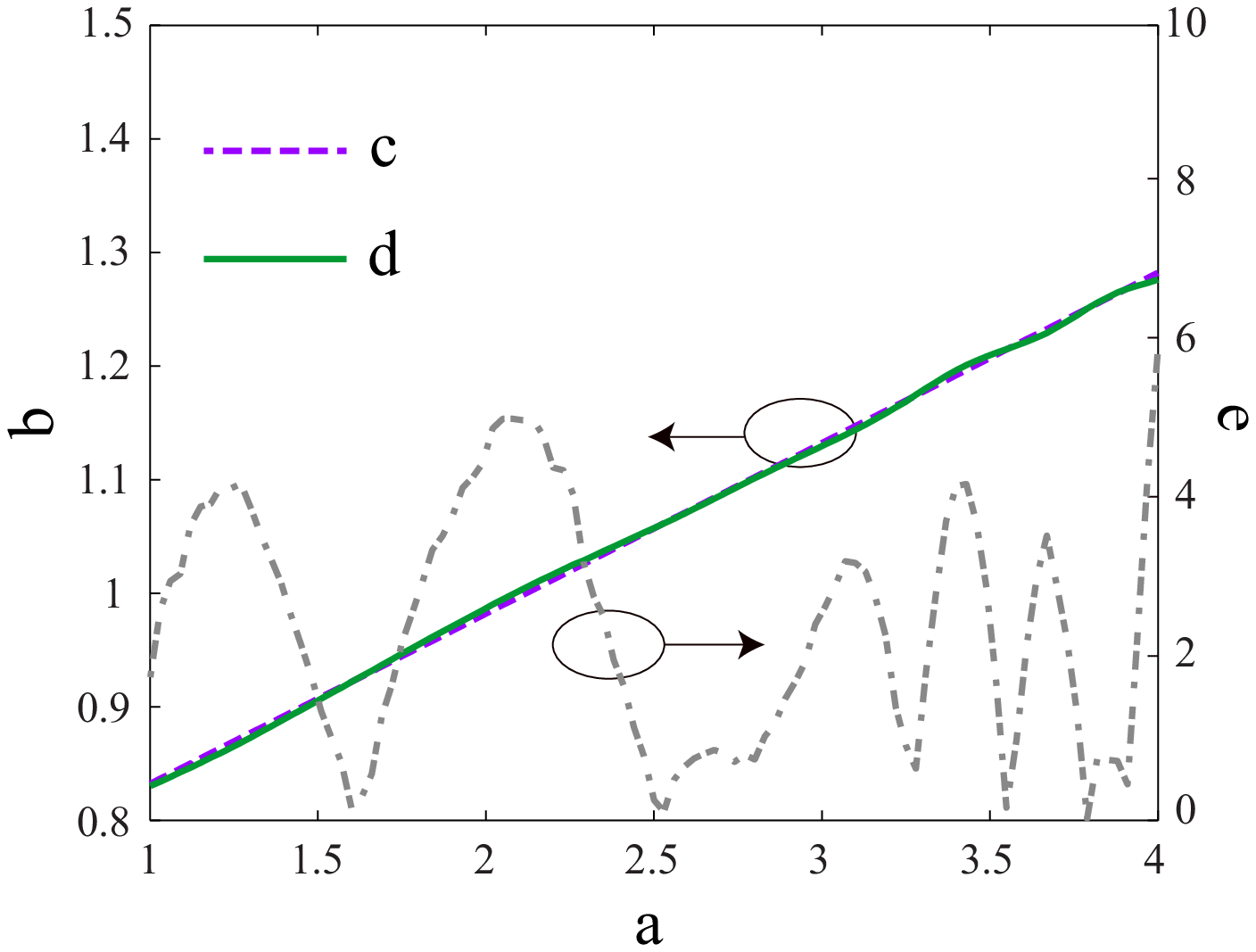}}
  \caption{(a) Specification is changed midway in the optimization example in Fig.~\ref{fig:pre} (starting from Fig.~\ref{fig:pre_2}); (b) fine model response aligning the new specification with the updated parameter $\boldsymbol{x}^{(3)}$; (c) fine model response aligning the new specification with the updated parameter $\boldsymbol{x}^{(4)}$.}
  \label{fig:mid}
\end{figure*}

Let us still use the above example to illustrate this point. Assume that at a point one obtained the response in Fig.~\ref{fig:pre_2}, corresponding to the $k$'s given in the second row of Tab.~II. Assuming $k_\text{fab}^\text{max}=0.24$, one sees that \mbox{$k^\text{max}=0.27>k_\text{fab}^\text{max}$}, meaning the synthesized DDS cannot be fabricated. Therefore, one needs to decrease the targeted swing, from $0.5$~ns to, say, $0.45$~ns, as shown in Fig.~\ref{fig:mid_1}. Note that the new fine model response is far from the new specification, although it closely followed the original specification. Then one employs the predistortion space mapping technique for the new specification. The updated parameter set $\boldsymbol{x}^{(3)}$ is given in the first row of Tab.~\ref{tab:mid} and the computed fine model response is plotted in Fig.~\ref{fig:mid_2}. Seeing that $k^\text{max}=0.234$ is smaller than $k_\text{fab}^\text{max}$, one keeps $0.45$~ns as the specified group delay swing in the following iterations.
One also observes that the new response follows the new specification, but that the error has not decreased so largely as that in Fig.~\ref{fig:pre_2}. This indicates that the variation in cross coupling is large. To speed up the convergence, one fixes then some parameters related to  cross coupling in $\boldsymbol{x}^{(3)}$ in the next iteration. Since the distance between C-sections $d_0$ is already fixed, one may further fix $f'_{02}$, $f'_{03}$, $f'_{04}$, $f'_{05}$, $f'_{06}$ and $f'_{07}$. Then one optimizes the other parameters in $\boldsymbol{x}^{(3)}$ at the new iteration. The new parameter set, $\boldsymbol{x}^{(4)}$, is given in the second row of Tab.~\ref{tab:mid}, and the computed fine model response is plotted in Fig.~\ref{fig:mid_3}. Note that the response very closely fits the specification and that the error has now receded below the allowed level. This example has shown that the predistortion space mapping technique provides more efficiency and flexibility than ASM, where the specification cannot be changed in the course of the procedure since $\boldsymbol{x}_c^*$ in \eqref{eq:f_error} is fixed in order to ensure convergence.

\begin{table*}[!t]
\renewcommand{\arraystretch}{2}
\caption{Computed fine model parameter sets at different iterations with a new specification ($f'_{0n}=\frac{\omega'_{0n}}{2\pi}/$GHz).}
\label{tab:mid}
\centering
\begin{tabular*}{1\textwidth}{@{\extracolsep{\fill}}c | c c c c c c c c c c c c c c}
\hline
\hline
$\boldsymbol{x}$ & $f'_{01}$ & $f'_{02}$ & $f'_{03}$ & $f'_{04}$ & $f'_{05}$ & $f'_{06}$ & $f'_{07}$ & $k'_1$ & $k'_2$ & $k'_3$ & $k'_4$ & $k'_5$ & $k'_6$ & $k'_7$\\
\hline
$\boldsymbol{x}^{(3)}$ & $2.025$ & $4.350$ & $4.350$ & $4.350$ & $4.350$ & $4.350$ & $4.350$ & $0.176$ & $0.234$ & $0.234$ & $0.234$ & $0.234$& $0.234$ & $0.234$\\
\hline
$\boldsymbol{x}^{(4)}$ & $1.976$ & $4.350$ & $4.350$ & $4.350$ & $4.350$ & $4.350$ & $4.350$ & $0.161$ & $0.222$ & $0.222$ & $0.222$ & $0.222$& $0.222$ & $0.222$\\
\hline
\hline
\end{tabular*}
\end{table*}

It is also interesting to realize that one can impose the convergence condition~\eqref{eq:cross_zero}, by fixing some parameters related to the cross coupling, even at a point of the procedure where the response was not diverging (i.e. where \eqref{eq:covergence_exam} is still satisfied). This indicates that~\eqref{eq:cross_zero} may be used as a strategy to speed up the convergence of~\eqref{eq:pre_iteration}.

\subsection{Experimental Verification}

To further validate the results of the full-wave simulation, we fabricated a prototype corresponding to the response in Fig.~\ref{fig:pre_3}.

\begin{figure}[!t]
\centering
  \psfrag{a}[c][c]{\footnotesize Frequency (GHz)}
  \psfrag{b}[c][c]{\footnotesize Group delay (ns)}
  \psfrag{c}[l][c]{\footnotesize Specified}
  \psfrag{d}[l][c]{\footnotesize Full-wave (MoM)}
  \psfrag{e}[l][c]{\footnotesize Measured}
  \psfrag{i}[l][c][0.7]{33~\text{mil}}
  \psfrag{j}[c][c][0.7]{19~\text{mil}}
  \psfrag{k}[c][c][0.7]{13~\text{mil}}
  \psfrag{g}[c][c][0.7]{\textcolor[rgb]{1.00,0.00,1.00}{CPW-stripline transition}}
  \psfrag{h}[c][c][0.7]{\textcolor[rgb]{0.44,0.00,0.87}{inner view}}
  \includegraphics[width=8cm]{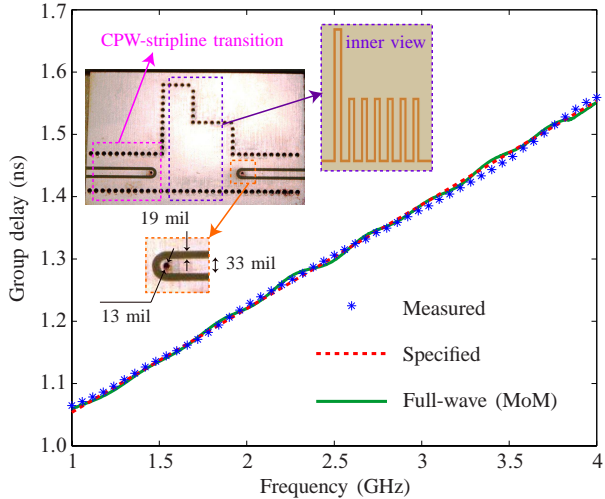}\\
  \caption{Measured group delay response of the fabricated prototype.}\label{fig:meas_gd}
\end{figure}

\begin{figure}[!t]
\centering
  \psfrag{a}[c][c]{\footnotesize Frequency (GHz)}
  \psfrag{b}[c][c]{\footnotesize $|S_{21}|$, $|S_{11}|$ (dB)}
  \psfrag{d}[l][c]{\footnotesize Full-wave (MoM)}
  \psfrag{e}[l][c]{\footnotesize Measured}
\psfrag{h}[c][c]{\footnotesize $|S_{21}|$}
\psfrag{k}[c][c]{\footnotesize $|S_{11}|$}
  \includegraphics[width=8cm]{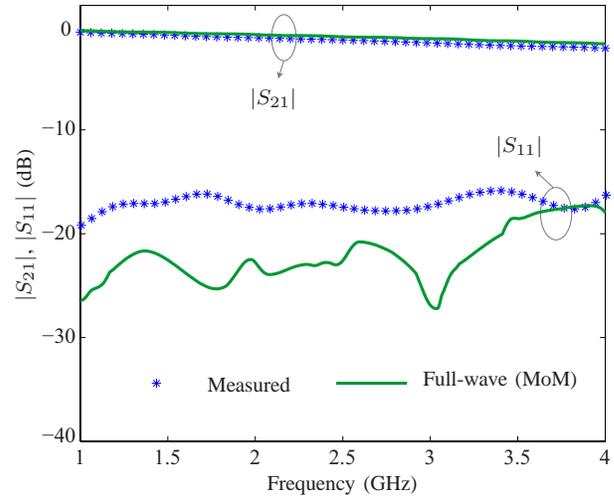}\\
  \caption{Measured transmission and reflection magnitude of the fabricated prototype.}\label{fig:meas_sc}
\end{figure}

The fabricated prototype is shown in the inset of Fig.~\ref{fig:meas_gd} and its dimension is shown in Tab.~\ref{tab:meas}. It is implemented in stripline technology and uses two via-hole based stripline to coplanar waveguide transitions for coplanar excitation, selected for measurement convenience. Via holes are placed around the structure to ensure a uniform voltage between the two ground planes of the stripline structure. The fabricated prototype is measured using a vector network analyzer and the experimental group delay and magnitude responses are plotted in Fig.~\ref{fig:meas_gd} and Fig.~\ref{fig:meas_sc}, respectively. Note that the measured group delay response closely follows the full-wave response and the specification, while the measured magnitudes slightly differ from the full-wave ones due to fabrication tolerance. The measured insertion loss has a maximum of $2.1$~dB at $4$~GHz, whereas the full-wave insertion loss is $1.7$~dB.

\begin{table}[!t]
\renewcommand{\arraystretch}{1.5}
\renewcommand{\tabcolsep}{1.5mm}
\caption{Dimensions of the fabricated prototype (unit: mil).}
\label{tab:meas}
\centering
\begin{tabular*}{0.48\textwidth}{@{\extracolsep{\fill}}c | c c c c c c c}
\hline
\hline
$n$ & $1$ & $2$ & $3$ & $4$ & $5$ & $6$ & $7$\\
\hline
$w_n$ & $7.20$ & $6.70$ & $6.70$ & $6.70$ & $6.70$ & $6.70$ & $6.70$\\
\hline
$s_n$ & $13.95$ & $9.17$ & $9.17$ & $9.17$ & $9.17$ & $9.17$ & $9.17$\\
\hline
$l_n$ & $436.35$ & $206.00$ & $206.00$ & $206.00$ & $206.00$ & $206.00$ & $206.00$\\
\hline
\hline
\end{tabular*}
\end{table}

\section{Conclusions}

SM techniques have been applied for the first time to the design of DDSs for analog signal precessing. Specifically, they have been applied to DDSs formed by coupled C-sections. Both conventional ASM and a new predistortion space mapping technique were applied and compared. The results indicate that the predistortion space mapping technique features enhanced accuracy and implementation simplification. Two full-wave and one experimental examples have been provided to illustrate the predistortion space mapping technique. From the authors' experience, predistorsion space mapping reduces the design of a stripline coupled C-section DDS from a dozen hours to about one hour using an FEM simulator accounting for all losses and for the presence of a film of epoxy adhesive.

\appendix

In this appendix, we shall present two examples of cascaded C-section DDS designs with loosely bounded coupling in the coarse model optimization.

\begin{figure}[!t]
  \center
  \psfrag{a}[c][c]{\footnotesize Frequency (GHz)}
  \psfrag{b}[c][c]{\footnotesize Group delay (ns)}
  \psfrag{e}[c][c]{\footnotesize Error ($10^{-4}$ ns)}
  \psfrag{c}[l][c]{\footnotesize $R_\text{spec}$}
  \psfrag{d}[l][c]{\footnotesize $R_c(\boldsymbol{x}_c^*)$}
  \includegraphics[width=8cm]{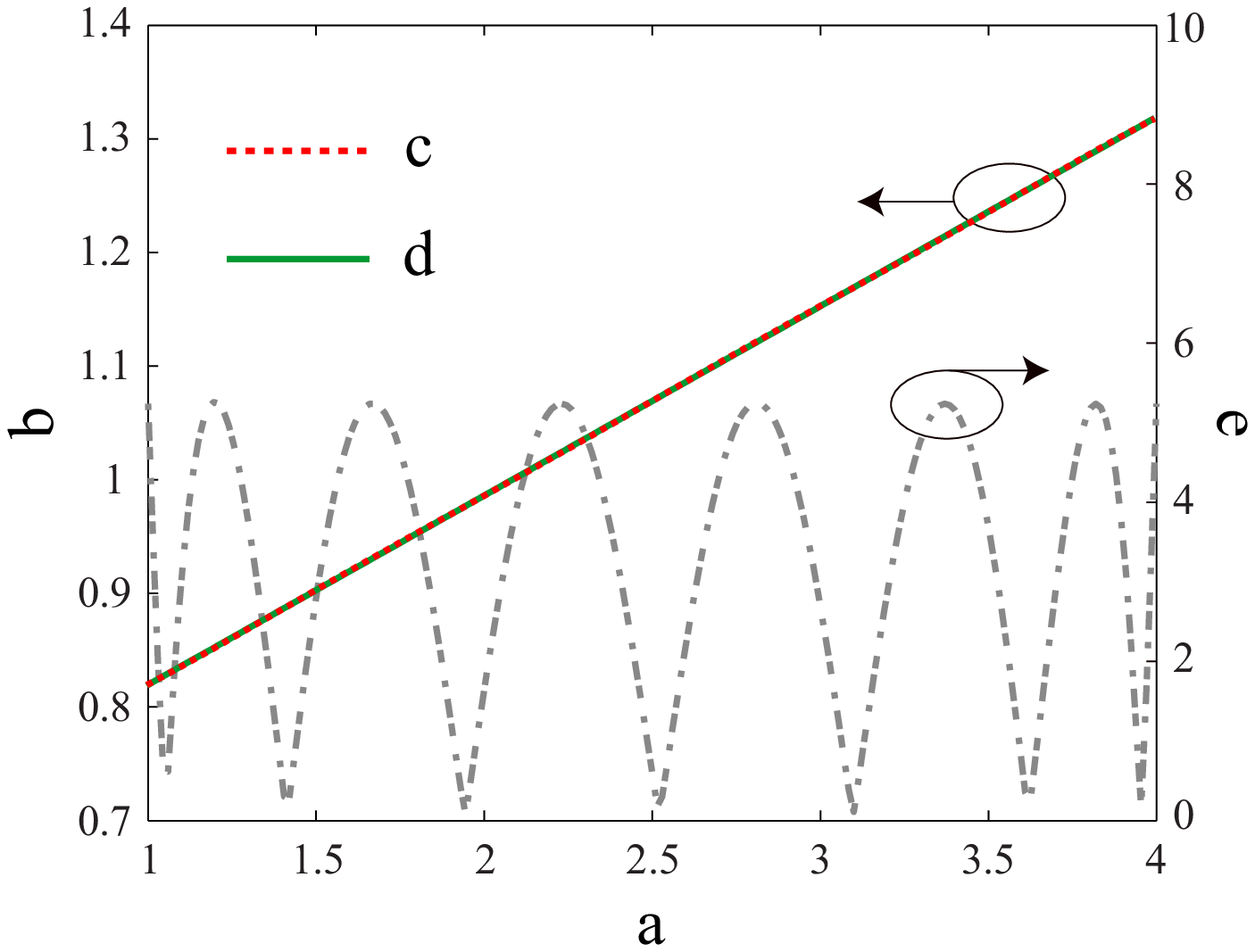}
  \caption{Calculated response and error for the coarse model with the parameter set in the first row of Tab.~\ref{tab:appen}.}
  \label{fig:appen1}
\end{figure}

In the first example, we set $U=0.5$ instead of $U=0.38$ used in Fig.~\ref{fig:asm_1}. The corresponding optimized parameter set $\boldsymbol{x}_c^*$ is shown in the first row of Tab.~\ref{tab:appen} and the response is plotted in Fig.~\ref{fig:appen1}. Note that three coupling coefficients are now different from each other while the rest four are identical. This larger degree of difference between the parameters is due to the fact that the coupling coefficients are now less saturated and therefore experience more freedom to change. Also note from Fig.~\ref{fig:appen1} that the error is reduced by a factor of about $10$ compared with Fig.~\ref{fig:asm_1}.

\begin{figure}[!t]
  \center
  \psfrag{a}[c][c]{\footnotesize Frequency (GHz)}
  \psfrag{b}[c][c]{\footnotesize Group delay (ns)}
  \psfrag{e}[c][c]{\footnotesize Error ($10^{-5}$ ns)}
  \psfrag{c}[l][c]{\footnotesize $R_\text{spec}$}
  \psfrag{d}[l][c]{\footnotesize $R_c(\boldsymbol{x}_c^*)$}
  \includegraphics[width=8cm]{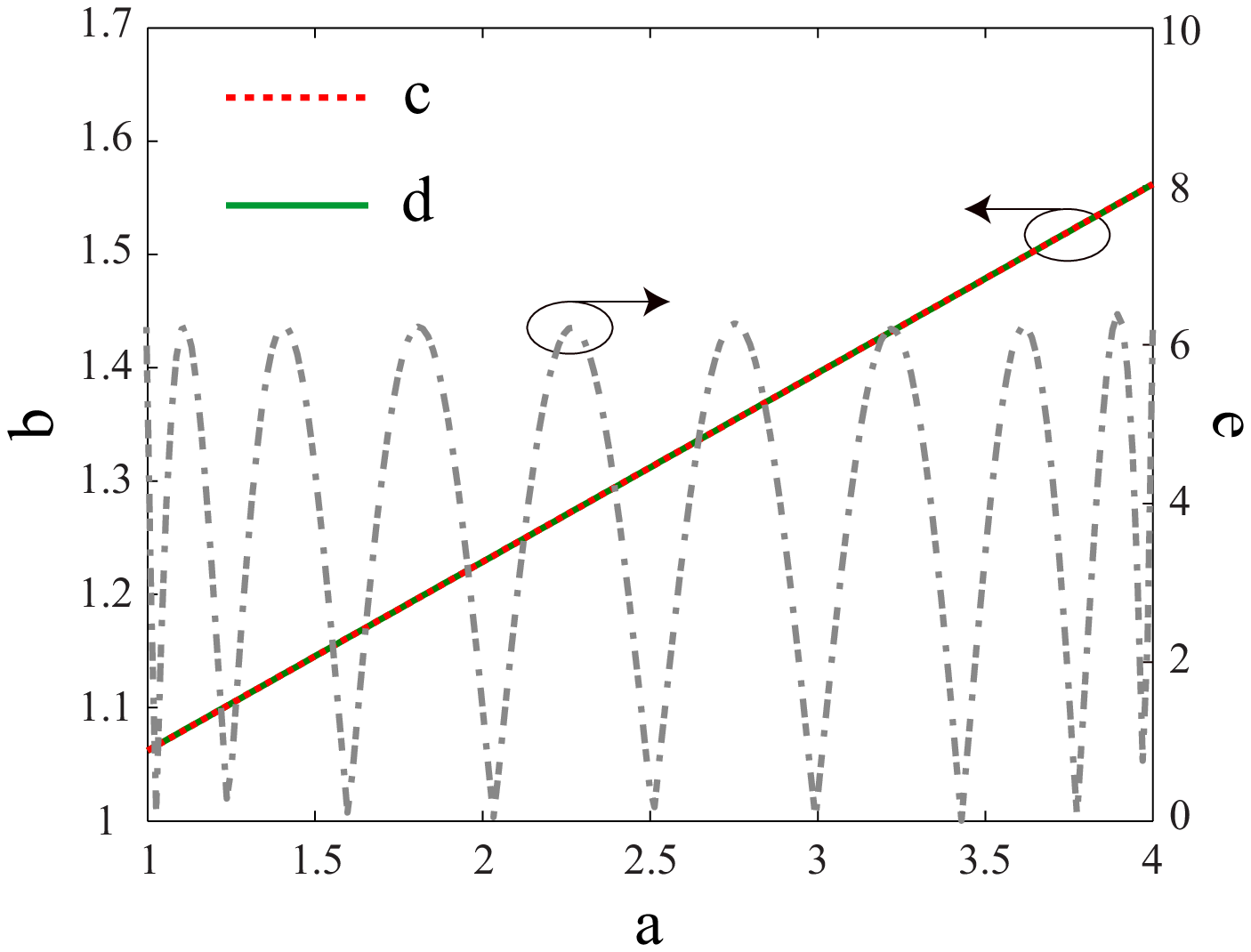}
  \caption{Calculated response and error for the coarse model with the parameter set in the second row of Tab.~\ref{tab:appen}.}
  \label{fig:appen2}
\end{figure}

In the second example, we further increase the upper bound to $U=0.7$. The optimized parameter set $\boldsymbol{x}_c^*$ is shown in the second row of Tab.~\ref{tab:appen} and the response is plotted in Fig.~\ref{fig:appen2}. Note that all the parameters are different from each other. Also note from Fig.~\ref{fig:appen1} that the error has now been reduced by a factor of about $100$ compared with Fig.~\ref{fig:asm_1}.

Although the error can be improved by increasing the upper bound of the coupling coefficients, as shown above, excessive bounds are inappropriate in practical applications, for two practical reasons. First, very high coupling coefficients are difficult to implement due to fabrication limitations. Second, the extremely small errors require extremely accurate coupling which might be impossible to achieve in the given fabrication tolerances.

\begin{table*}[!t]
\renewcommand{\arraystretch}{2}
\caption{Optimized coarse parameter set $\boldsymbol{x}_c^*$ with different upper bounds ($f_{0n}=\frac{\omega_{0n}}{2\pi}/$GHz).}
\label{tab:appen}
\centering
\begin{tabular*}{1\textwidth}{@{\extracolsep{\fill}}c | c c c c c c c c c c c c c c}
\hline
\hline
$U$ & $f_{01}$ & $f_{02}$ & $f_{03}$ & $f_{04}$ & $f_{05}$ & $f_{06}$ & $f_{07}$ & $k_1$ & $k_2$ & $k_3$ & $k_4$ & $k_5$ & $k_6$ & $k_7$\\
\hline
$0.5$ & $1.592$ & $2.813$ & $3.852$ & $4.845$ & $4.845$ & $4.845$ & $4.845$ & $0.094$ & $0.353$ & $0.399$ & $0.500$ & $0.500$& $0.500$ & $0.500$\\
\hline
$0.7$ & $0.160$ & $4.110$ & $4.575$ & $2.816$ & $2.125$& $4.834$ & $4.840$ & $0.034$ & $0.404$ & $0.616$ & $0.309$ & $0.195$ & $0.683$ & $0.690$ \\
\hline
\hline
\end{tabular*}
\end{table*}

\bibliographystyle{IEEEtran}
\bibliography{mybib}

\begin{biography}[{\includegraphics[width=1in,height=1.25in,clip,keepaspectratio]{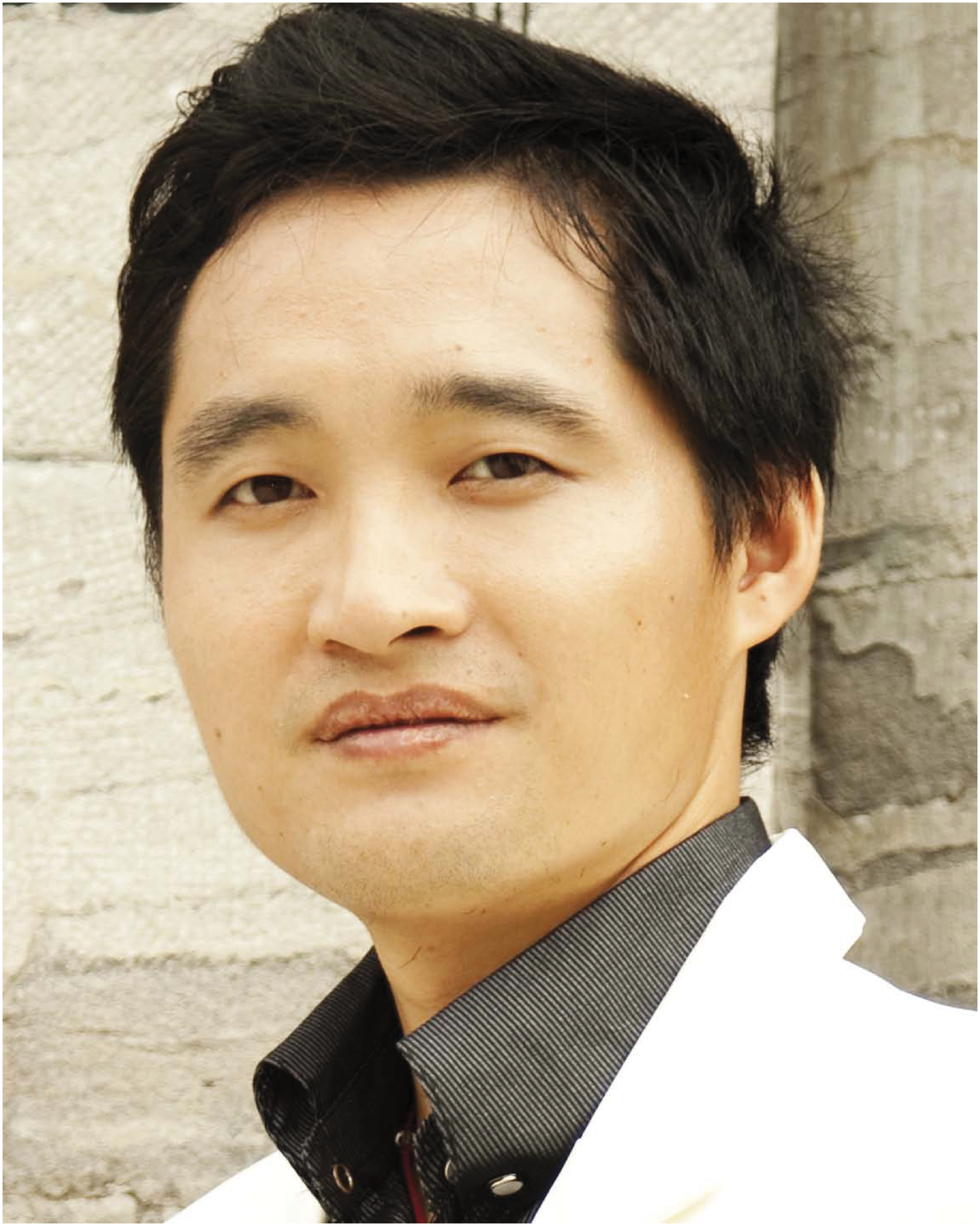}}]
{Qingfeng Zhang} (S'07-M'11) was born in Changzhou, China. He received the B.E. degree from University of Science and Technology of China (USTC), Hefei, China, in 2007, and the Ph.D. degree from Nanyang Technological University, Singapore, in 2011, all in electrical engineering.

He joined Poly-Grames Research Center of \'{E}cole Polytechnique de Montr\'{e}al, Montr\'{e}al, Canada, as a postdoctoral fellow in 2011. He has authored and co-authored 21 technical journal papers. His current research interests include synthesis theory of filters and dispersive delay structures, real-time radio systems, array antennas and leaky-wave antennas, meta-material, magnet-less non-reciprocal systems, space mapping and other optimization techniques.
\end{biography}

\begin{biography}[{\includegraphics[width=1in,height=1.25in,clip,keepaspectratio]{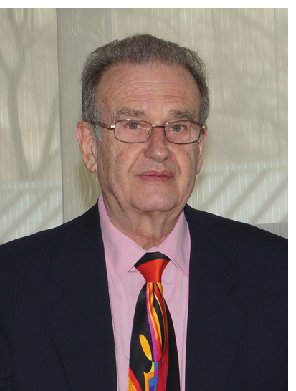}}]
{John W. Bandler} (S'66-M'66-SM'74-F'78-LF'06) studied at the Imperial College of Science and Technology, London, England, and received the B.Sc. (Eng.), Ph.D., and D.Sc. (Eng.) degrees from the University of London, London, England, in 1963, 1967, and 1976, respectively.

He joined McMaster University, Hamilton, ON, Canada, in 1969. Currently, he is a Professor Emeritus. He was President of Optimization Systems Associates Inc. (OSA), which he founded in 1983, until 1997, when OSA was acquired by Hewlett-Packard Co. He is President of Bandler Corporation, Dundas, ON, Canada, which he founded in 1997. He has served on editorial boards, technical program review committees, as guest editor of several special issues, and has published more than 480 technical papers.

Dr. Bandler is a Fellow of several societies, including the Canadian Academy of Engineering and the Royal Society of Canada (since 1986). He received the IEEE MTT S 2004 Microwave Application Award ``For application of optimization technology, design with tolerances and yield-driven design to microwave devices, circuits and systems." He received the IEEE Canada 2012 A.G.L. McNaughton Gold Medal, which honors ``outstanding Canadian engineers recognized for their important contributions to the engineering profession." He was also honored in 2012 by a Queen Elizabeth II Diamond Jubilee Medal. For his lifetime achievements in the field of microwave theory and techniques, he received the IEEE MTT-S 2013 Microwave Career Award.

Dr Bandler has contributed to commercial high-frequency computer-aided design tools, design with tolerances, yield-driven design, high-fidelity electromagnetics-based optimization, and the art, science and cognitive interpretation of space mapping, a technology he introduced in 1994, and which explains the engineer's mysterious ``feel" for a problem.

\end{biography}

\begin{biography}[{\includegraphics[width=1in,height=1.25in,clip,keepaspectratio]{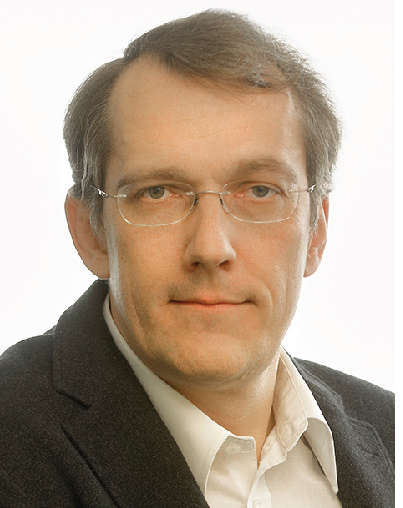}}]
{Christophe Caloz} (S'00-A'00-M'03-SM'06-F'10) received the Dipl\^{o}me d'Ing\'{e}nieur en \'{E}lectricit\'{e} and the Ph.D. degree from \'{E}cole Polytechnique F\'{e}d\'{e}rale de Lausanne (EPFL), Switzerland, in 1995 and 2000, respectively.

From 2001 to 2004, he was a Postdoctoral Research Engineer at the Microwave Electronics Laboratory of University of California at Los Angeles (UCLA). In June 2004, Dr. Caloz joined \'{E}cole Polytechnique de Montr\'{e}al, where he is now a Full Professor, the holder of a Canada Research Chair (CRC) and the head of the Electromagnetics Research Group. He has authored and co-authored over 500 technical conference, letter and journal papers, 12 books and book chapters, and he holds several patents. His works have generated about 10,000 citations.

Dr. Caloz is a Member of the Microwave Theory and Techniques Society (MTT-S) Technical Committees MTT-15 (Microwave Field Theory) and
MTT-25 (RF Nanotechnology), a Speaker of the MTT-15 Speaker Bureau, the Chair of the Commission D (Electronics and Photonics) of the Canadian Union de Radio Science Internationale (URSI) and an MTT-S representative at the IEEE Nanotechnology Council (NTC). In 2009, he co-founded the company ScisWave, which develops CRLH smart antenna solutions for WiFi. Dr. Caloz received several awards, including the UCLA Chancellor's Award for Post-doctoral Research in 2004, the MTT-S Outstanding Young Engineer Award in 2007, the E.W.R. Steacie Memorial Fellowship in 2013, the Prix Urgel-Archambault in 2013, and many best paper awards with his students at international conferences. He is an IEEE Fellow. His research interests include all fields of theoretical, computational and technological electromagnetics, with strong emphasis on emergent and multidisciplinary topics, including particularly metamaterials, nanoelectromagnetics, exotic antenna systems and real-time radio.

\end{biography}

\end{document}